\documentclass[english,twocolumn,showpacs,preprintnumbers,amsmath,amssymb]{revtex4}
\usepackage[T1]{fontenc}
\usepackage[latin1]{inputenc}
\usepackage{amsmath}
\usepackage{graphicx}

\makeatletter

\providecommand{\tabularnewline}{\\}

\makeatletter

\makeatletter

\usepackage{amsfonts}
\usepackage{amssymb}

\usepackage{yfonts}

\usepackage{dcolumn}

\makeatother

\makeatother

\usepackage{babel}
\makeatother
\begin{document}

\title{Thermodynamics of pairing in mesoscopic systems}

\author{Tony Sumaryada}

\author{Alexander Volya}

\affiliation{Department of Physics, Florida State University, Tallahassee, FL
32306-4350, USA}

\date{\today}

\begin{abstract}
Using numerical and analytical methods implemented for different models
we conduct a systematic study of thermodynamic properties of pairing
correlation in mesoscopic nuclear systems. Various quantities are
calculated and analyzed using the exact solution of pairing. An in-depth
comparison of canonical, grand canonical, and microcanonical ensemble
is conducted. The nature of the pairing phase transition in a small
system is of a particular interest. We discuss the onset of discontinuity
in the thermodynamic variables, fluctuations, and evolution of zeros
of the canonical and grand canonical partition functions in the complex
plane. The behavior of the Invariant Correlational Entropy is also
studied in the transitional region of interest. The change in the
character of the phase transition due to the presence of magnetic
field is discussed along with studies of superconducting thermodynamics. 
\end{abstract}



\maketitle

\section{Introduction\label{sec::intro}}

Pairing correlations and related superconducting or superfluid properties
are robust features of quantum many-body systems. In physics anywhere
from quarks to stars it is hard to find systems that under certain
conditions do not exhibit pairing correlations. The Cooper phenomenon
\cite{cooper56}, namely the instability against formation of particle-pairs
in a macroscopic Fermi-system under an arbitrarily weak attractive
force, is a primary reason for thriving of pairing.

Pairing in mesoscopic systems, such as atomic nuclei \cite{Dean:2002zx},
metal clusters \cite{Borrmann:2001,Braun:1998,Braun:1999}, ultra
small grains \cite{vonDelft:2001}, quantum dots \cite{Harting:2000},
interacting spins \cite{KAWABATA:1970,SUZUKI:1969}, has attracted
a lot of attention recently. Indeed, questions of phase transitions
\cite{Belic:2003nz,Borrmann:2000,Chen:1990,Chomaz:2005,cooper56},
interplay with other collective modes \cite{Bahri:1998,Barranco:1999},
continuum effects \cite{Barranco:2001} and thermodynamical properties
of small systems are important for the present day science and technology.

In this work we conduct a systematic study of thermodynamics of pairing
correlations in small systems. We use two-types of model Hamiltonians
of lower and higher symmetry where the pairing problem is solved exactly
and all quantum states are identified. We use a quasi-spin algebra
with the effective numerical implementation to obtain a full solution
for systems ranging in size from a few particles to as large as over
a hundred of particles. The traditional BCS solution is also considered
for comparison. Using these results we compare different thermodynamic
ensembles: microcanonical, canonical and grand canonical. The differences
indicate a mesoscopic nature of the system \cite{Schiller:2001,Schiller:2002,Schiller:2003}
and diminish in the macroscopic limit. Some discrepancies observed
in thermodynamics are related to non-thermal nature of the pure pairing
interaction \cite{RefWorks:900} and raise questions of equilibration
and thermalization. Through thermodynamic ensembles and using invariant
correlational entropy we study and analyze the pairing phase transition
as a function of temperature or excitation energy, magnetic field,
size of the system, and pairing strength. 

We further explore the evolution of zeros in the complex temperature
plane for the canonical ensemble \cite{BESTGEN:1969,GROSSMAN-ROSENHAU:1969,GROSSMAN.S:1968,GROSSMAN.S:1969},
where recent findings established clear correlations of pair breaking
with peaks in entropy and branches of complex temperature roots approaching
real axis \cite{Schiller:2002,Ipsen:2003,Dean:2002zx}. We extend
this discussion with consideration of the phase transition based on
the Yang-Lee theory \cite{Bena:2005,LEE:1952,YANG:1952}. The study
of the system in the magnetic field, evolution of zeros in the partition
function as a function of the field strength, spin fluctuations and
the change of the phase transition type are particularly interesting.

The presentation below is structured as follows: we first introduce
the pairing Hamiltonian, identify properties of the pairing problem
and define models for our study in Sec. \ref{sec:Pairing-Hamiltonian}.
In Sec. \ref{sec:BCS} we consider a BCS approximation which shows
the generic features of a paired system. The bulk of the work is presented
in Sec. \ref{sec:Statistical-treatment} and its subsections, where
different methods are introduced, discussed and compared. In Sec.
\ref{sec:Magnetic-Properties} we concentrate on the effects that
external magnetic field or rotation have on the properties of paired
systems; this includes the classification of the phase transitions
using the distribution of zeros in the partition functions.

\section{Pairing Hamiltonian\label{sec:Pairing-Hamiltonian}}

We approach the pairing problem by defining a pair of two single-particle
states denoted here as ${1}$ and $\tilde{1}$. This pair-wise identification
can be based on an arbitrary symmetry; however, the fundamental symmetry
with respect to time reversal is the most common. For this work we
assume a pair as two particles in time-conjugated single-particle
states that due to this symmetry have identical energies. Using the
language of the second quantization the pair creation and annihilation
operators are $p_{1}^{\dagger}=a_{{1}}^{\dagger}a_{\tilde{1}}^{\dagger}$
and $p_{1}=a_{\tilde{1}}a_{{1}}$, respectively. Here the $a_{1}^{\dagger}$
and $a_{1}$ are single-particle creation and annihilation operators
with the usual fermion commutation rules. The pair is labeled by the
same single-particle index 1, and is invariant under the time conjugation,
$p_{1}=p_{\tilde{1}}$, since $a_{\tilde{\tilde{1}}}=a.$

The algebra of the pair operators on a pair-state 1 (a pair of orbitals
1 and $\tilde{1}$) is identical to that of an SU(2) spin algebra
called quasi-spin, in general the commutation relations are \begin{equation}
[p_{1}^{\dagger},\, p_{2}]=2\delta_{12}\, p_{1}^{z},\end{equation}
 where \begin{equation}
p_{1}^{z}=\left(n_{1}-\frac{1}{2}\right)\,,\end{equation}
the operator related to the particle number $n_{1}=(a_{{1}}^{\dagger}a_{{1}}+a_{\tilde{1}}^{\dagger}a_{\tilde{1}})/2$
operator for the pair-state ${1}.$

A pair-state $(1,\tilde{1})$ occupied by a pair or completely empty
correspond to quasi-spin $1/2$ with projections $p_{1}^{z}=1/2$
and $p_{1}^{z}=-1/2$, respectively. Alternatively, these states are
referred to as states with seniority $s_{1}=0$ identifying the number
of unpaired nucleons in the pair-state $1$. The states with one unpaired
particle correspond to $s_{1}=1$ and to zero quasi-spin.

The most general form of the two-body Hamiltonian that describes motion
of pairs at fixed particle number is \begin{equation}
H=2\sum_{1>0}\epsilon_{1}n_{1}-\sum_{1,2>0}{G_{12}}p_{1}^{\dagger}p_{2},\label{ham}\end{equation}
 where the summation runs over pair-orbitals, denoted as $1>0$, $\epsilon_{1}$
are single-particle energies, and $G_{12}=G_{21}$ determines the
strength of pair scattering. Using the quasi-spin the same Hamiltonian
can be written as \begin{equation}
H=\sum_{1>0}\epsilon_{1}+\sum_{1>0}2\left(\epsilon_{1}-\frac{G_{11}}{2}\right)p_{1}^{z}-\sum_{12>0}{G_{12}}\,\left({\vec{p}}_{1}\cdot{\vec{p}}_{2}-p_{1}^{z}p_{2}^{z}\right).\label{qph}\end{equation}

The problem is analogous to the Heisenberg model of $\Omega/2-s$
interacting spins $|{\vec{p}}|=1/2\,,$ with the Zeeman splitting
created by the single-particle energies. The $\Omega/2$ stands here
for the total number of double-degenerate levels and $s=\sum_{1}s_{1}$
represents the total seniority. Because of the magnetic-field like
splitting the total quasi-spin vector ${\vec{p}}=\sum_{1>0}{\vec{p}}_{1}$
is not conserved, while the remaining cylindrical symmetry allows
for the conservation of the $z$-projection $p^{z}=N/2-\Omega/4\,,$equivalent
to the total particle number $N=2\sum_{1>0}n_{1}$.

The eigenstates of the Hamiltonian (\ref{ham}) are identified by
the set of $\Omega/2$ seniorities ${\bf s}=\{ s_{1}\}$ denoting
the available and blocked pair-states. In the language of the spin
model (\ref{qph}) seniorities represent the number of spin 1/2 particles
in the system, thus totally removing all blocked states from interaction.
The Hamiltonian within a certain seniority partition ${\bf s}$ is
given as \begin{eqnarray}
H_{{\bf s}}=\sum_{1>0}s_{1}\epsilon_{1}+2\sum_{1>0}^{{\bf s}}n_{1}\left(\epsilon_{1}-\frac{G_{11}}{2}\right)\\
-\sum_{1\ne2}^{{\bf s}}{G_{12}}\,\left({\vec{p}}_{1}\cdot{\vec{p}}_{2}-p_{1}^{z}p_{2}^{z}\right) &  & ,\nonumber \end{eqnarray}
 where the upper summation limit ${\bf s}$ implies that all blocked
states with $s_{1}=1$ are excluded.

Since each unpaired particle doubles the degeneracy of the many-body
state the total degeneracy of a given eigenstate is $g_{{\bf s}}=2^{s}$.
With other symmetries, beyond the time reversal, the degeneracy of
states can be higher. Additional degeneracies such as the one due
to the rotational symmetry can further reduce the problem to larger
values of the quasi-spin. In the spherical shell model within a given
$j$-shell there are total of $\omega_{j}=j+1/2$ time-conjugate pair
states, and the total quasi-spin is preserved by the pairing interaction.
For such $j$-shell a quasi-spin vector ${\vec{p}}_{j}=\frac{1}{2}\,\sum_{m}{\vec{p}}_{jm}$
can be introduced which together with the number operator for this
level and its own hermitian conjugate again forms an SU(2) group.
The independence of matrix elements and quasi-spin operators on magnetic
sub-states allows to rewrite the Hamiltonian \eqref{ham} as \begin{equation}
H=\sum_{j}\epsilon_{j}N_{j}-\sum_{jj'}V_{jj'}P_{j}^{\dagger}P_{j'},\label{eq:HamiltonianSym}\end{equation}
 where for the reasons of the two-particle state normalization a pair
operator and interaction matrix elements are redefined as follows
\begin{equation}
\vec{P}_{j}=\frac{1}{2\sqrt{\omega_{j}}}\,\sum_{m}{\vec{p}}_{jm}\,,\label{eq:Gj}\end{equation}
\[
V_{jj'}=\sqrt{\omega_{j}\omega_{j'}}G_{jj'}.\]
 The exact diagonalization of the pairing Hamiltonian (\ref{ham})
or \eqref{eq:HamiltonianSym}, depending on the symmetries of the
model, is performed using the quasi-spin algebra. The ability to obtain
all many-body states with a relatively simple exact treatment of pairing
is an important component in this study. The more detailed discussion
of the seniority based diagonalization can be found in Ref. \cite{Auerbach:1966,RefWorks:773,Dukelsky:2001}.
We refer to the exact treatment of pairing as EP. The applications
of algebraic methods extend far beyond our models; treatments of proton-neutron
pairing as well as more exotic forms of pairing-type Hamiltonians
are discussed in Ref. \cite{Pang:1969,Pang:1968,Hecht:1965,Hecht:1989,Pang:1967,Dussel:1986,Evans:1981,Engel:1998,Ginocchi.Jn:1965}.
Other methods of exact solution, analogies with boson-fermion models
and electrostatic analogies should be mentioned \cite{Richardson:1964,Richards.Rw:1965,Dukelsky:2006,Dukelsky:2002,FengPan:1999,FengPan:2002}
.

Below in Sec. \eqref{sec:Statistical-treatment} we introduce thermodynamic
ensembles and discuss thermodynamic variables used to study the many-body
system that undergoes pairing phase transition. For each of the cases
we construct the partition function exactly based on the full numerical
solution to the pairing problem. As our examples we consider two basic
types of systems. The picket-fence (or ladder system) which has $\Omega/2$
equally spaced double-degenerate levels, where the total fermion capacity
is $\Omega$. The level spacing is chosen as the unit of energy. The
picket-fence model is a minimal symmetry system with the time reversal
only; therefore the degeneracy of each eigenstate $\alpha$ is $g_{\alpha{\bf s}}=2^{s}.$
A second model with only two levels, but of large degeneracy, represents
an opposite {}``high symmetry'' case. Spacing between the two levels
is again taken as the unit of energy. Due to additional symmetry,
the degeneracy of many-body states is higher. The action of the pairing
Hamiltonian is limited to either diagonal or level to level pair transfer.
For the two-level system with an appropriate selection of the basis
states the Hamiltonian matrix is tri-diagonal. This facilitates substantially
the numerical treatment, making determination of all many-body states
in systems with a hundred or more particles possible. The two types
of model spaces with total occupancy $\Omega,$ the particle number
$N,$ and the constant pairing strength $G$ constitute the set of
input parameters in this study. Introduction of the magnetic field
in Sec.\ref{sec:Magnetic-Properties} does not require a separate
diagonalization, however requires determination of the total spin
projections onto an axis parallel to the direction of the field. We
note that the total number of many-body states is ${\bf \Omega}=\frac{\Omega!}{N!(\Omega-N)!}.$

\section{BCS\label{sec:BCS} }

The BCS approximation is the common approach to tackle the pairing
problem. While this method is asymptotically exact in thermodynamic
limit it still produces remarkably good results for smaller systems.
The BCS method assumes the presence of a condensate and approximates
the dynamics of interacting particles \eqref{ham} with a motion of
independent quasi-particles. Although most of the issues that we intend
to address in this work can not be fully explored within the BCS picture
due to its limitations, the method is a good benchmark for many of
the questions and an excellent guidance to the dynamical regions of
interest. Below we review the approach while stressing some of the
key elements relevant to this work.

Within the BCS theory the general pairing Hamiltonian in Eq. (\ref{ham})
is brought to an approximate single particle form using Bogoliubov's
transformation. The parameters of the transformation, the set of gaps
$\Delta_{1}$and chemical potential $\mu,$ are determined via gap
equations \begin{equation}
\Delta_{1}=\frac{1}{2}\sum_{2>0}\, G_{12}\frac{\Delta_{2}}{e_{2}},\label{eq:BCSGAP}\end{equation}
 and the chemical potential is given by the particle number \begin{equation}
N=2\sum_{1>0}n_{1}\quad{\rm where}\quad n_{1}=\frac{1}{2}(1-\frac{\varepsilon_{1}}{e_{1}}).\label{eq:BCSmu}\end{equation}
 For simplicity of notations we introduce single particle energies
shifted by the chemical potential and the diagonal interaction strength
$\varepsilon_{1}=\epsilon_{1}-\mu-G_{11}/2.$ The result of the Bogoliubov
transformation is the spectrum of states given by the independent
quasi-particle excitations with energies \begin{equation}
e_{1}=\sqrt{\varepsilon_{1}^{2}+\Delta_{1}^{2}}.\label{eq:quasiparticle}\end{equation}
 The total energy of the paired system is \[
E=2\sum_{1>0}\left(\epsilon_{1}-\frac{G_{11}}{2}\right)n_{1}-\sum_{1,2>0}G_{12}\frac{\Delta_{1}\Delta_{2}}{4e_{1}e_{2}}.\]
 As earlier, the summations here go over the pair-states.

In this work for all our models we use a constant pairing strength
$G_{11}\equiv G$ which due to Eq. (\ref{eq:BCSGAP}) leads to a constant
pairing gap for all single particle pairs, $\Delta_{1}\equiv\Delta.$
A single parameter for the interaction strength, in our view, allows
for the most transparent study of the important features, the results
are generic, and the methods of BCS and EP are applicable to general
situations. For constant pairing the BCS gap equation and the energy
are a textbook examples: \begin{equation}
1=\frac{G}{2}\sum_{1>0}\frac{1}{e_{1}},\quad E=2\sum_{1>0}\left(\epsilon_{1}-\frac{G}{2}\right)n_{1}-\frac{\Delta^{2}}{G}.\label{eq:BCS-constG}\end{equation}
 To accommodate the cases with higher symmetry following Eq. \eqref{eq:Gj}
it is convenient to introduce $V=\omega G$ where $\omega$ is the
pair degeneracy which is level independent in both picket-fence ($\omega_{j}=1$)
and two-level ($\omega_{j_{1}}=\omega_{j_{2}}\equiv\omega)$ models.

The particle number non-conservation intrinsic to the Bogoliubov transformation
is one of the problems associated with the BCS applications to mesoscopic
systems. Furthermore, in a system with discrete levels Eq. (\ref{eq:BCSGAP})
may not have a solution, with the exception of a trivial case $\Delta_{1}=0$.
Formally, this transitional point \cite{Belyaev:1959} corresponds
to the critical interaction strength where the largest eigenvalue
of the matrix built from the elements $(G_{12}\varepsilon_{1}+G_{21}\varepsilon_{2})/(4\varepsilon_{1}\varepsilon_{2})$
is equal to unity. The interpretation of this is that at a low pairing
strength the pairing is too weak to overcome gaps in the single particle
spectrum which leads to a normal state. This situation is again specific
to small systems where it appears in contrast to the Cooper instability
\cite{cooper56}. The total absence of the pairing correlations below
the critical pairing strength is a second major drawback of the BCS
approach in mesoscopic systems. Exact solutions indicate a gradual
dissipation of pairing correlations extending almost to zero strength
\cite{Belyaev:1959,Zelevinsky:1996,RefWorks:742,Dean:2002zx}. The
critical pairing strength as determined by the BCS is still an important
parameter identifying the location of the mesoscopic phase transition.

An analytic solution to the BCS equations can be obtained for the
system of two levels defined above. For a half-occupied system the
chemical potential due to the particle-hole symmetry is an exact average
of the monopole-renormalized single-particle energies $\mu=(\epsilon_{1}+\epsilon_{2}-G)/2.$
Thus, \begin{equation}
\Delta^{2}=V^{2}-\left(\frac{\Delta\epsilon}{2}\right)^{2}\,,\quad\Delta\epsilon=\epsilon_{1}-\epsilon_{2}.\label{eq:BCS-2l}\end{equation}
 The introduction of the renormalized strength $V$ makes this equation
independent of $\Omega$.

\begin{figure}[ht]
\includegraphics[width=8cm]{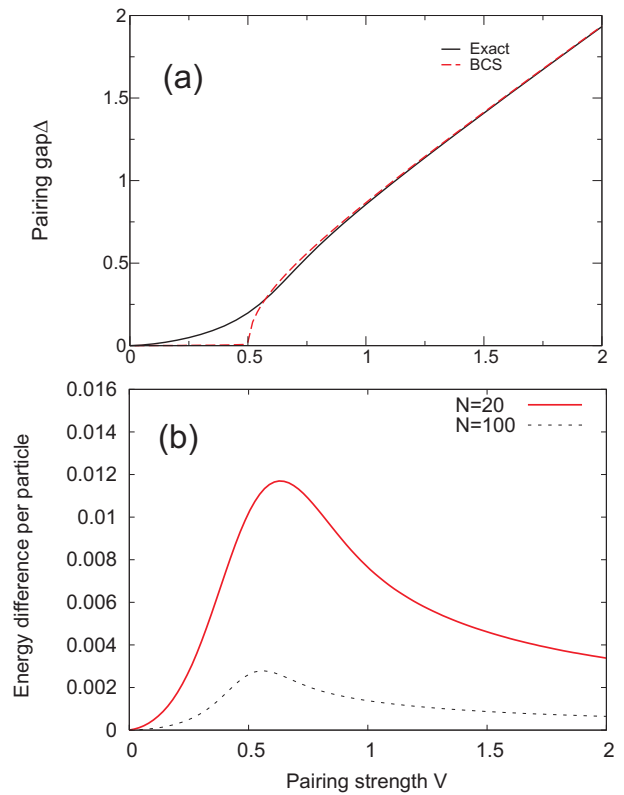}

\caption{(Color online) In the upper panel the BCS pairing gap is shown as
a function of the pairing strength for the two-level, half-occupied
system with 20 particles. In the lower panel the energy difference
per particle between BCS and the exact result is shown as a function
of the pairing strength for the same $N$=20 system and is compared
with the results for a larger half-occupied two-level model containing
100 particles. \label{delgn20}}
\end{figure}

In Fig. \ref{delgn20}(a) the BCS gap is plotted as a function of
energy for a two-level model following Eq. (\ref{eq:BCS-2l}). The
curve has a square-root discontinuity at the critical pairing strength
$V_{cr}$=0.50 in the units of level spacing. The concept of the gap
does not appear in the exact solution, however this quantity can be
deduced from the energy associated with paring correlations. The second
curve in the same figure shows the gap computed through Eq. (\ref{eq:BCS-constG})
where energy and occupation numbers are obtained from the exact solution.
The difference between these two curves depicts the shortcoming of
the BCS when applied to a small system; for related discussions and
comparison of BCS with exact techniques see Refs. \cite{Pradhan:1973,Burglin:1996,Dukelsky:2003}.
In Fig. \ref{delgn20}(b) an alternative view on the EP-BCS comparison
is given. Here we show the energy difference per particle between
BCS and EP as a function of the pairing strength for N=20 and 100
particles. As the particle number grows the BCS and EP become equivalent.
The peak in the BCS-EP discrepancy appears in the pairing phase transition
region, around $V_{cr}\approx0.6$ which is close to an analytically
obtained BCS value of 0.5. The discrepancy in $V_{cr}$ is known to
arise from the pair-vibrations and other renormalizations of the BCS
ground state \cite{Broglia:2000,RefWorks:764}.

For our second (picket-fence) model, the critical pairing strength
can be determined in the case of a half-occupied system with an even
number of levels through the sum of a harmonic series \begin{equation}
G_{cr}=\frac{\Delta\epsilon}{\left(\sum_{n=1}^{(\Omega/2+1)/2}\frac{1}{n}+\ln4\right)},\label{eq:BCS-ladd}\end{equation}
 which in the limit of a large number of levels converges to zero
logarithmically $G_{cr}\sim\Delta\epsilon/\ln\Omega$. We remind here
that formally for this model $G=V$ since $\omega_{j}=1$ and $\Omega/2$
equals to the number of levels. This logarithmic dependence in the
macroscopic limit is related to the exponential dependence of the
gap on the pairing strength and density of states near the Fermi surface,
which represents the Cooper instability. 

We conclude this section with a note on the BCS approach at finite
temperature $T=1/\beta$. By modeling the thermodynamics of quasiparticles
with non-interacting Fermi gas we obtain a modified version of the
Eq. \eqref{eq:BCS-constG} \begin{equation}
1=\frac{G}{2}\sum_{1>0}\frac{\tanh\left(\frac{\beta}{2}e_{1}\right)}{e_{1}},\label{eq:TBCSGap}\end{equation}
 where quasiparticle energies are of the form (\ref{eq:quasiparticle}).
A related discussion of thermodynamic treatment within the grand canonical
partition function and following from it thermal BCS is presented
in Sec. \ref{sub:Grand-Canonical-ensemble}.

For the two-level, half occupied model the temperature dependence
of the critical pairing strength is given by

\begin{equation}
V_{cr}=\frac{\Delta\epsilon}{2}\coth(\frac{\beta\Delta\epsilon}{4}).\label{eq:VBthermal}\end{equation}

\section{Statistical treatment\label{sec:Statistical-treatment}}

Statistical properties of a many-body systems are addressed using
normalized density operators \cite{Blum:1981}, usually referred to
as statistical operators $\hat{w}$ \cite{Landau:1978} and defined
as \begin{equation}
\hat{w}(E,N)=\frac{1}{Z}\,\delta(E-\hat{H})\delta(N-\hat{N})\label{eq:denmicro}\end{equation}
 for the microcanonical, \begin{equation}
\hat{w}(\beta,N)=\frac{1}{{\cal Z}}\,\exp(-\beta\hat{H})\,\delta(N-\hat{N})\label{eq:dencano}\end{equation}
 for the canonical, and \begin{equation}
\hat{w}(\beta,\mu)=\frac{1}{{\bf Z}}\,\exp\left(-\beta(\hat{H}-\mu\hat{N})\right)\label{eq:partitionf}\end{equation}
 for the grand canonical ensemble. In the above definitions the parameter
$\beta=1/T$ refers to an inverse temperature and $\mu$ corresponds
to the chemical potential. Here we use units where the Boltzmann constant
is equal to unity, allowing units of energy to be used for temperature.
The normalization constants $Z,\,\,{\cal Z}$, and ${\bf Z}$ are
the partition functions for the corresponding ensembles; so that the
statistical operators are normalized by the trace ${\rm Tr}(\hat{w})=1.$
The statistical averages are calculated as \begin{equation}
\langle\hat{O}\rangle={\rm Tr}(\hat{O}\hat{w}).\label{eq:operator}\end{equation}
 The entropy for the above ensembles is defined as \begin{equation}
S=-\langle\ln(\hat{w})\rangle=-{\rm Tr}(\hat{w}\,\ln\hat{w})\,.\label{eq:entropy:wn}\end{equation}
The above definition is strictly speaking applicable only for thermally
equilibrated system which makes thermodynamical Boltzmann-Gibbs entropy
discussed below equivalent to the von-Neumann entropy of a quantum
ensemble in Eq. (\ref{eq:entropy:wn}). A new light on complexity
of quantum states in non-thermalized or non-equilibrated systems can
be obtained with the invariant correlational entropy \cite{Sokolov:1998}
(ICE) that also appears to be a good tool to study the phase transitions
in mesoscopic systems \cite{Stoyanov:2004,Volya:2003su}. The correlational
entropy is defined through the behavior of the microcanonical density
matrix (\ref{eq:denmicro}) for each individual quantum state in response
to a noise in an external parameter. For the purposes of this work
we consider pairing strength $V$ to be this external parameter. The
variations in $V$ within the interval $[V,V+\delta V]$ result in
an averaged density operator

\[
\hat{\overline{w}_{\alpha}}=\frac{1}{\delta V}\,\int_{V}^{V+\delta V}\hat{w}_{\alpha}(V),\]
 where the weight operator $\hat{w}_{\alpha}$ is a density operator
for an individual quantum state $\alpha$ followed with evolution
of $V$, for a fixed parameter $V$ this is a projection operator.
The averaged statistical weight matrix is used to obtain the ICE via
Eq. (\ref{eq:entropy:wn}).

The quality or applicability of a given thermodynamic approach to
a small system is often under question. While in some studies various
ensembles are used interchangeably, there are significant dangers
on this path. Our investigations below not only show up the pairing
phase transition and its evolution as a function of the particle number
but also draw attention to some subtle differences in thermodynamic
treatments.

\subsection{Canonical ensemble}

Given an exact solution to the pairing problem via diagonalization
in the seniority scheme, Sec. \ref{sec:Pairing-Hamiltonian}, the
formal definition \eqref{eq:dencano} can be written explicitly for
the eigenstates labeled by $\alpha$ and ${\bf s}$: \begin{equation}
w_{\alpha{\bf s}}=\frac{1}{{\cal Z}}\,\exp(-\beta E_{\alpha{\bf s}})\,,\quad{\rm where}\quad\label{eq:CA:w}\end{equation}
 \begin{equation}
{\cal Z}(\beta,N)=\sum_{\alpha{\bf s}}g_{\alpha{\bf s}}\exp(-\beta E_{\alpha,{\bf s}})\label{eq:CA:Z}\end{equation}
is the canonical partition function. The ensemble average \eqref{eq:operator}
for any quantity is given as \begin{equation}
\langle O\rangle=\sum_{\alpha{\bf s}}g_{\alpha s}w_{\alpha s}\,\langle\alpha{\bf s}|O|\alpha{\bf s}\rangle,\label{eq:CA-average}\end{equation}
 where $\langle\alpha{\bf s}|O|\alpha{\bf s}\rangle$ is the quantum-mechanical
expectation value for the corresponding operator in the eigenstate
$\alpha$ with the seniority set ${\bf s}$. The entropy is given
via the usual expression \begin{equation}
S=-\sum_{\alpha{\bf s}}g_{\alpha s}w_{\alpha s}\ln(w_{\alpha s}).\label{eq:CA:S}\end{equation}
 The reader may be familiar with the following set of traditional
thermodynamic relations \cite{pathria} \begin{equation}
\langle E\rangle=-\frac{\partial}{\partial\beta}\ln({\cal Z}),\label{eq:dere}\end{equation}
 the entropy $S$ can be found directly from the statistical definition
\eqref{eq:entropy:wn} \begin{equation}
S=\ln{\cal Z}+\beta\langle E\rangle=-\frac{\partial F}{\partial T}.\end{equation}
 The Helmholtz free energy is defined as\begin{equation}
F=-T\ln({\cal Z})=\langle E\rangle-TS.\end{equation}
 The Eq. \eqref{eq:dere} involves a derivative, however in our calculations
we avoid numerical differentiations always going back to the definition
(\ref{eq:CA-average}). For example specific heat is computed using
its relation to the energy fluctuations $\langle(E-\langle E\rangle)^{2}\rangle$,
\begin{equation}
C=\left(\frac{\partial\langle E\rangle}{\partial T}\right)=\beta^{2}\frac{\partial^{2}\ln{\cal Z}}{\partial\beta^{2}}=\beta^{2}\langle(E-\langle E\rangle)^{2}\rangle.\end{equation}

\begin{figure}[ht]
\includegraphics[clip,width=8.5cm]{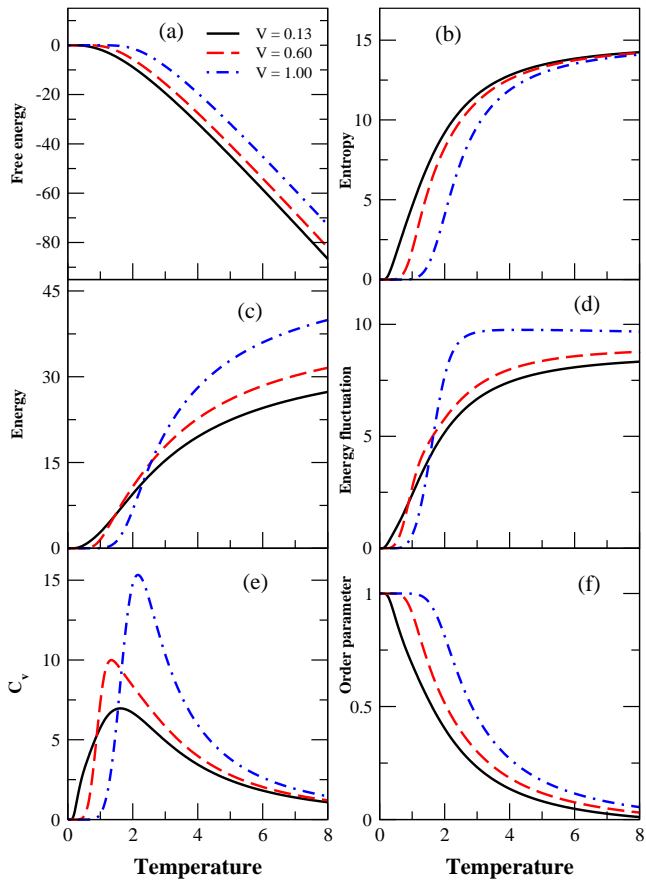}

\caption{(Color online) (a) Free energy, (b)Entropy, (c)Energy, (d) Energy
fluctuations, (e) Specific heat, and (f) Order parameter of a ladder
system with 12 levels and 12 particles as a function of temperature.
\label{can12L1}}
\end{figure}

\begin{figure}[ht]
\includegraphics[width=9cm]{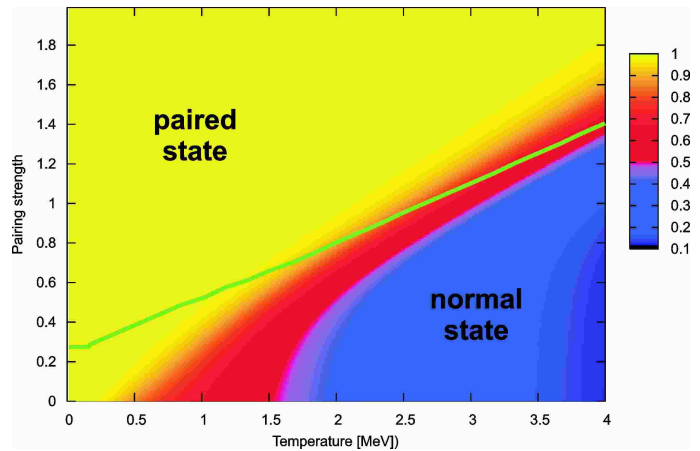}

\caption{(Color online) The contour plot of the order parameter as a function
of the pairing strength and temperature. Half occupied 12-level system
is shown. The line separates normal and paired regions based on the
BCS equation. \label{contourm}}
\end{figure}

\begin{figure}[ht]
\vskip 0.4 cm \includegraphics[clip,width=8.5cm]{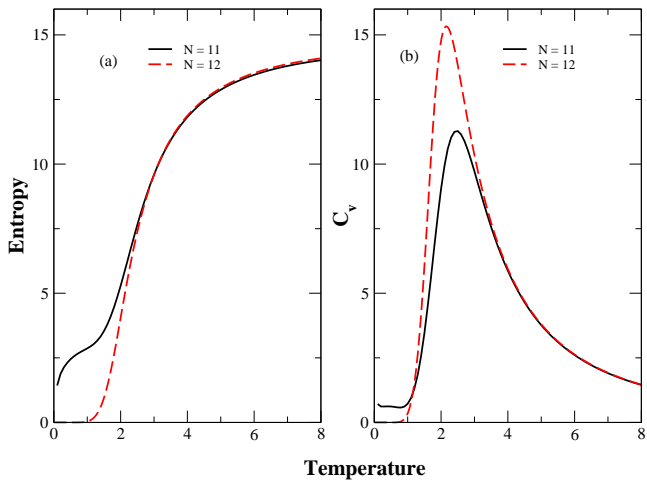}

\caption{(Color online) (a) Entropy and (b) Specific heat as a function of
temperature, for an odd and even number of particles and $V=1$.\label{can11}}
\end{figure}

\begin{figure}[ht]
\vskip 0.6 cm \includegraphics[clip,width=8.5cm]{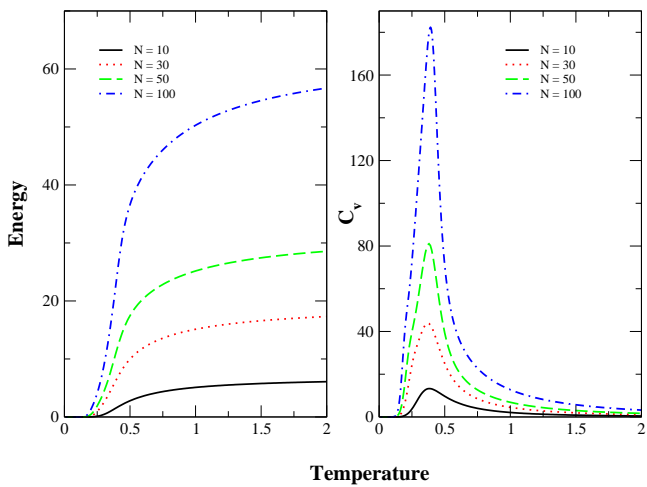}

\caption{(Color online) (a) Energy and (b) Specific heat as a function of
temperature, for $V=1$ and various number of particles N=10, 30,
50, and 100. \label{ECcan2L}}
\end{figure}

\begin{figure}
\includegraphics[width=5cm]{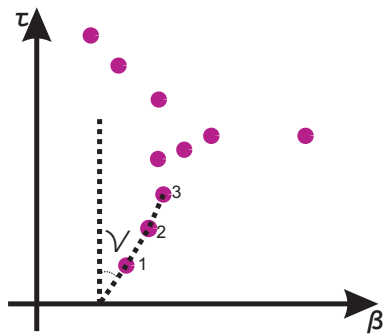}

\caption{(Color online) The lowest branch of zeros computed for N=100 and
G=1.00 is schematically shown. \label{fig:Evolution-of-DOZ} }
\end{figure}

\begin{figure*}
\includegraphics[clip,width=15cm]{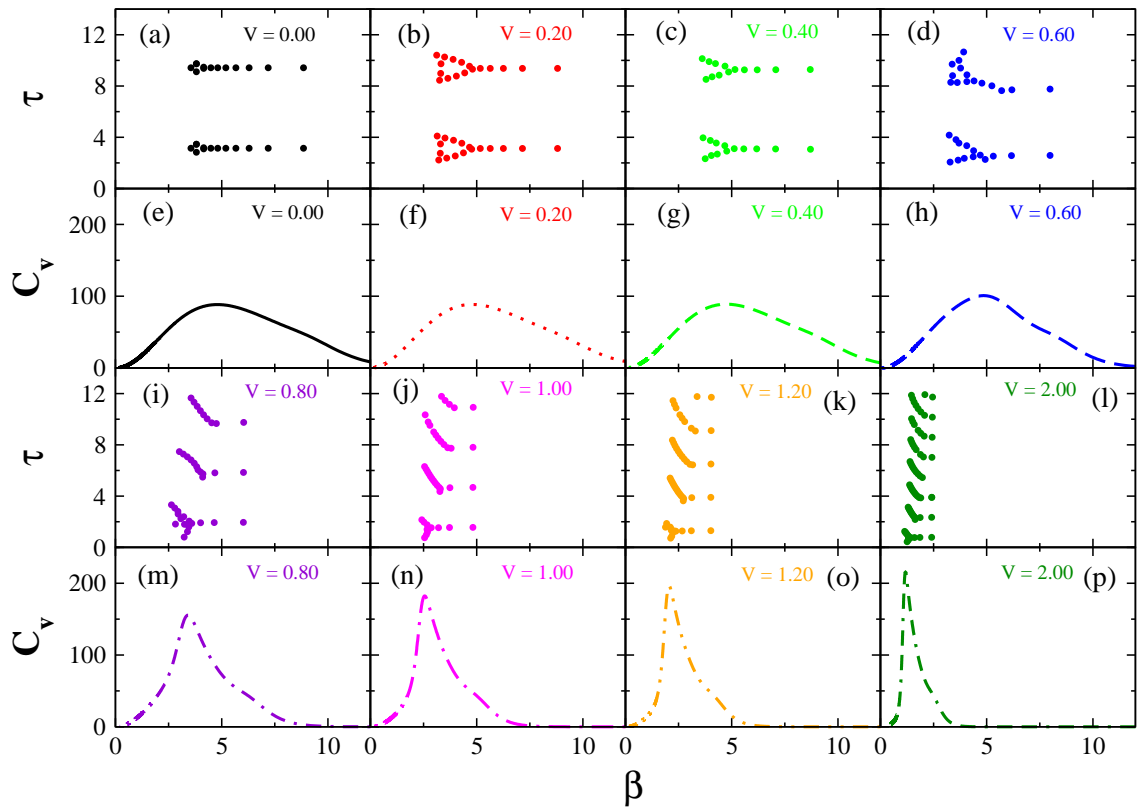}

\caption{(Color online) Evolution of DOZ and~$C_{v}$~in the complex temperature
plane $\mathcal{B}=\beta+i\tau$ for $N$=100 particles in the half-occupied
two-level system. There are number of poles near and exactly on the
imaginary axis, they are of no interest to our discussion and are
not shown. The poles for other systems are discussed in \cite{Dean:2002zx}
and references therein, the interpretation and the nature of branches
is discussed in \cite{Schiller:2002av}, further in-depth exploration
of the above model can be found in Ref. \cite{Ipsen:2003}. \label{dozcv}}
\end{figure*}

\begin{figure}
\includegraphics[clip,width=8.5cm]{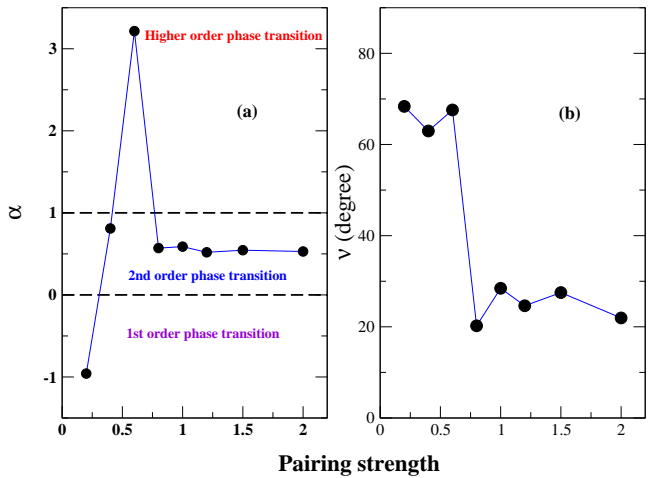}

\caption{(Color online) Parameter of the phase transition for the two level
system with $N$=100 particles as a function of pairing strength (a)
$\alpha$~vs $V$ and (b) transition angle $\nu$~ vs $V$. \label{fig:PTtype}}
\end{figure}

The results of our study based on the canonical ensemble are shown
in Fig. \ref{can12L1}-\ref{fig:PTtype}. In Fig. \ref{can12L1} (a
- d) free energy, entropy, energy, and energy fluctuation of the ladder
system with 12 levels and 12 particles are shown as a function of
temperature, similar study may be found in \cite{Dean:2002zx,Guttormsen:2001}
and references therein. The critical pairing strength for this model
from BCS, Eq. (\ref{eq:BCS-ladd}), at zero temperature is $V_{cr}=0.27$.
The curves correspond to different pairing strengths showing various
conditions: weak pairing with about half the critical pairing strength
$V=0.13$; pairing strength above the critical value $V=0.6$; and
strong pairing $V=1.$ All of the plots show essentially similar trends:
there is a sharp change in each of the quantities as a function of
temperature in a certain region. This region is associated with the
phase transition from the paired to the normal state. Most transparently
it can be seen in \ref{can12L1}(e) where it is associated with the
peak in heat capacity. The critical temperature $T_{cr}$ depends
on the pairing strength. It can be observed that the transitional
region for strong pairing ($V>V_{cr}$ for $T=0$) is roughly consistent
with the BCS, which gives $T_{cr}=2.7$ and 1.3 for $V=1$ and 0.6,
respectively. Naturally, the stronger pairing interactions support
the superconducting state at higher temperature or excitation energy.
For weak pairing the transitional behavior is present at zero temperature.
This is consistent with the earlier finding that pairing correlations
appear in the ground state even for small $V$. The decline of weak
pairing ($V\le0.13$) phase is still associated with the peak in heat
capacity which becomes smaller as the pairing strength is weakened,
while staying essentially at the same $T_{cr}\sim1.3.$ 

The phase diagram can be further explored by considering an order
parameter which we define here as a fraction of paired particles $\psi=(N-\langle s\rangle)/N$,
the $\langle s\rangle$ is the ensemble-averaged value of the total
seniority. The dependence of the order parameter on temperature, shown
in Fig. \ref{can12L1}(f), shows that the fraction of superconducting
pairs drops sharply in the transitional region which is also identified
by the critical behavior of other thermodynamic quantities.

The contour plot of the order parameter as a function of the pairing
strength and temperature is shown in Fig. \ref{contourm}. The shaded
area in the upper left corresponds to the high percentage of particles
in the condensate, which occurs at low temperature and high pairing
strength; while in the opposite limit the superconducting state disappears.
The solid line indicates the phase boundary as follows from the BCS
approximation. We note that at zero temperature the fraction of superconducting
particles is high even at zero pairing strength this special point
corresponds to the absence of two-body interactions which results
in pair-wise Fermi occupation of time-reversed orbitals.

Throughout this work we mainly discuss systems with an even particle
number; we found that the difference between odd and even systems
in the critical region of interest is small. Most of the distinction
occurs at zero temperature where degeneracy of an odd-particle ground
state and non-zero spin are important. This can be seeing in Fig.
\ref{can11} where we compare the entropy and specific heat as a function
of temperature for $N=11$ and $N=12$ 12-level ladder systems. 

The transition to the thermodynamic limit is explored for a two-level
system in Fig. \ref{ECcan2L}. Unless noted otherwise, in our study
we select exactly half-occupied systems with $N=\Omega/2$. The region
of interest is identified by the peak in heat capacity seen in Fig.
\ref{ECcan2L}(b). With the increased particle number this peak becomes
sharper as expected in the macroscopic limit, where the phase transition
is represented by a discontinuity. Another interesting remark can
be made about the location of the peak. Following Eq. (\ref{eq:VBthermal})
within the BCS approximation the location of the phase transition
for a half-occupied two-level model does not depend on size of the
valence space $\Omega,$ at $V=1$ the BCS prediction is $T_{c}=0.455.$
As seen from the figure this is not exactly correct, for a small 10-particle
system the peak appears at about $T_{c}=0.35,$ and only with the
increase in the particle number the peak moves right to the BCS predicted
value, thus confirming the BCS as an exact theory in the macroscopic
limit. 

In recent years analysis of poles in the complex temperature plane
and the evolution of branches of these poles has attracted a lot of
attention as a study and classification tool for mesoscopic phase
transitions. The theory related to the distribution of zeros (DOZ)
in fugacity of the grand canonical ensemble dates back to Yang-Lee
\cite{LEE:1952,YANG:1952}. Later works \cite{GROSSMAN.S:1968,GROSSMAN.S:1969,GROSSMAN-ROSENHAU:1969}
extended it the to the complex temperature plane of the canonical
ensemble. The method of classifications of mesoscopic phase transitions,
recently suggested in Ref. \cite{Borrmann:2000} is based on the distribution
of zeros near the real axis. Some of the interesting questions such
as whether the nature of the phase transition changes as a function
of size have been studied with this approach. The first steps in the
analysis of mesoscopic systems undergoing pairing phase transitions
were done in Ref. \cite{Schiller:2002av,Dean:2002zx}, the evolution
of DOZ and comparison with the thermal BCS for a two-level model can
be found in Ref. \cite{Ipsen:2003}.

In what follows we use the classification of phase transitions developed
by Bormann et.al. \cite{Borrmann:2000}. We introduce complex temperature
as $\mathcal{B}=\beta+i\tau$ and numerically seek a set of zeros
$\mathcal{B}_{i}$ in the canonical partition function ${\cal Z}(\mathcal{B}_{i},N)=0$,
since the function is real the zeros appear in complex conjugate pairs
and we can limit the region of consideration to $\tau\ge0$. The product
expansion of the partition function in terms of zeros using the Weierstrass
theorem gives \begin{equation}
{\cal Z}(\mathcal{B})={\bf \Omega}\,\prod_{i}\left(1-\frac{\mathcal{B}}{\mathcal{B}_{i}}\right)\left(1-\frac{\mathcal{B}}{\mathcal{B}_{i}^{*}}\right).\label{eq:weierstrass}\end{equation}
 The DOZ in the complex temperature plane for the two-level system
is shown schematically in Fig.\ref{fig:Evolution-of-DOZ}.%

The sets of zeros form branches \cite{Schiller:2002,Ipsen:2003},
in Fig.\ref{fig:Evolution-of-DOZ} only the branch lowest to the real
axis is shown. The size of the system determines the distance between
neighboring zeros which in macroscopic limit becomes continuous. Phase
transitions are associated with branches crossing the real axis. Indeed
the zeros in the partition function appear as poles in thermodynamic
variables; for energy or heat capacity we have from (\ref{eq:weierstrass})
\begin{equation}
\langle E(\mathcal{B})\rangle=\sum_{i}\left(\frac{1}{\mathcal{B}_{i}-\beta}+\frac{1}{\mathcal{B}_{i}^{*}-\beta}\right),\label{eq:EB}\end{equation}
 \begin{equation}
C_{V}=\beta^{2}\sum_{i}\left(\frac{1}{(\mathcal{B}_{i}-\beta)^{2}}+\frac{1}{(\mathcal{B}_{i}^{*}-\beta)^{2}}\right).\label{eq:CV}\end{equation}
In general, although there are no poles at the real axis, the derivative
$d^{k}(\ln Z)/d\beta^{k}\sim\sum_{j}(\mathcal{B}_{i}-\beta)^{-k}$
may result in a divergent sum. As suggested in \cite{Borrmann:2000}
the classification of phase transitions in the Ehrenfest sense can
be extended to a smaller system by considering how the discrete roots
of the phase transition branch approach the real axis. By labeling
the roots in the phase transition branch starting from the closest
one to real axis, see Fig. \ref{fig:Evolution-of-DOZ}, the crossing
angle can be given as \[
\nu=\arctan\frac{\beta_{2}-\beta_{1}}{\tau_{2}-\tau_{1}}.\]
 The power law that expresses the congestion of roots as they approach
real axis at $\tau\rightarrow0$ determines the second parameter $\alpha$
as $|\mathcal{B}_{i+1}-\mathcal{B}_{i}|\sim\tau_{i}^{-\alpha}$.

The first order phase transition, which in thermodynamic limit appears
as a discontinuity in the first derivative of the free energy corresponds
to a vertical uniform approach of poles $\nu=0,\,\:\alpha=0.$ In
other cases the transition is of the second order for $0<\alpha<1$
or of a higher order if $\alpha>1.$ This classification establishes
a condition at which poles in sums of the form \eqref{eq:EB} and
\eqref{eq:CV} accumulate a logarithmically divergent series. For
a vertical approach, $\nu=0$ at the critical temperature the $|\mathcal{B}_{j}-\beta_{cr}|\sim j^{1/(\alpha+1)}$
therefore $k$-th derivative of the partition function would lead
to a divergent series if $k\ge\alpha+1$.

To find poles in the complex plane we developed a numerical technique
that uses analiticity of the above functions. We first determine the
number of roots in a desired region using a contour integral\begin{equation}
n=\frac{1}{2\pi i}\oint\langle E(\mathcal{B})\rangle d\mathcal{B}.\label{eq:nroots}\end{equation}
 The line integration is fast and is done avoiding paths that go directly
over the roots, this assures numerical stability and the real and
integer result of Eq. \eqref{eq:nroots} guarantees the accuracy.
Once the number of roots is known we use a method in the spirit of
the Laguerre's polynomial root finding technique \cite{Press:1992}
. The problem is mathematically analogous to the two-dimensional problem
of electrostatics. In the numerical method we converge to a given
{}``charge'' in the presence of the field from other {}``charges''
which is modeled via multipole expansion using the analytically known
derivatives of the {}``field strength''. The found roots are sequentially
removed, namely balanced by the {}``charge'' of an opposite sign.
Depending on the starting point and the density of roots, the numerical
cancellation is not always perfect, and the same root may appear several
times. Given that the total number of roots is known this problem
is easily fixed by choosing a different starting point or by exploring
a smaller region. In the calculations we stabilize the sum in the
partition function by selecting scaling so that the largest term in
the sum \eqref{eq:CA:Z} equals unity.

A series of plots where evolution of poles in the complex temperature
plane as a function of the pairing strength is shown in Fig. \ref{dozcv}.
The behavior of the heat capacity as a function of temperature for
each case is shown below to highlight the phase transition point.
With no pairing, $V=0$, the zeros are distributed along the two (almost)
horizontal lines. Similar picture is seen at the pairing strength
significantly below critical ($V_{cr}=0.5$ at zero temperature from
BCS). At about the critical strength, $V=0.4$, a noticeable bifurcation
occurs with the lower branch evolving toward the real axis. As pairing
strength increases, the branches move down and more branches becomes
visible in our figures; in Fig. \ref{dozcv} we use the same temperature
scale for all values of $V$. The lowest branch that approaches the
real axis is associated with the phase transition. The latter is confirmed
by the peak in the heat capacity that becomes sharper in cases with
stronger pairing.

In Fig. \ref{fig:PTtype} the dependence of the critical parameters
$\nu$ and $\alpha$ on the pairing strength is addressed. Below the
critical pairing strength the curves fluctuate, here, there is no
phase transition and $\nu$ and $\alpha$ can not be interpreted as
critical parameters. At a sufficiently strong pairing interaction,
however, the behavior of the parameters stabilizes showing a second
order phase transition.

\subsection{Grand Canonical ensemble\label{sub:Grand-Canonical-ensemble}}

The grand canonical ensemble is of a special importance in statistical
mechanics, since the partition function for non-interacting particles,
${\bf Z}_{0}(\beta,\,\mu),$ is given by an analytical expression.
The grand canonical partition function can be determined using the
canonical one, \begin{equation}
{\bf Z}(\beta,\,\mu)=\sum_{N=0}^{\Omega}z^{N}\,{\cal Z}(\beta,\, N),\label{eq:GCPart2}\end{equation}
 where fugacity $z=\exp(\beta\mu)$ is introduced. For non-interacting
Fermi particles \[
{\bf Z}_{0}(\beta,\,\mu)=\prod_{1}\left(1+z\,\exp(-\beta\epsilon_{1})\right),\]
 where $\epsilon_{1}$ is a single-particle spectrum. The above expression
results, for example, in the commonly used form for the occupation
numbers \[
n_{i}=\left(1+\exp[\beta(\epsilon_{i}-\mu)]\right)^{-1}.\]
 The grand canonical approach and the use of the above Fermi distribution
for small systems with a fixed number of particles is common, however
may present serious problems. On the other hand, even for non-interacting
particles computation of the microcanonical or canonical partition
function is difficult \cite{Pratt:2000}. Investigation of the mesoscopic
limit where statistical ensembles may no longer be equivalent is of
interest here.

The grand canonical ensemble is advantageous even when it comes to
interacting systems, the partition function can be expressed via diagrammatic
summation. In relation to pairing we mention here a method first proposed
in Ref. \cite{Gaudin:1960}, a more detailed discussion can be found
in \cite{Langer:1964}. The full partition function for constant (or
factorizable) pairing interaction can be obtained as\begin{equation}
{\bf Z}={\bf Z}_{0}\,\int_{0}^{\infty}dt\,\exp(-Y(t)),\label{eq:GCPartition}\end{equation}
 where function $Y(t)$ is \[
Y(t)=t-4\sum_{1>0}\ln\left[\frac{\cosh\left(\frac{\beta}{2}\sqrt{\epsilon_{1}^{2}+\frac{Gt}{2\beta}}\right)}{\cosh\left(\frac{\beta\epsilon_{1}}{2}\right)}\right].\]
 The most straightforward saddle point approximation to the integral
(\ref{eq:GCPartition}) leads to a saddle point $t_{s}$, which we
express in terms of a gap parameter as $\Delta^{2}=Gt_{s}/(2\beta).$
Thus, the saddle-point equation becomes a familiar gap equation of
thermal BCS \eqref{eq:TBCSGap}, and the thermal BCS theory represents
the lowest order approximation of the grand canonical expression in
Eq. \eqref{eq:GCPartition}.

\begin{figure}[ht]
\includegraphics[clip,width=8.5cm]{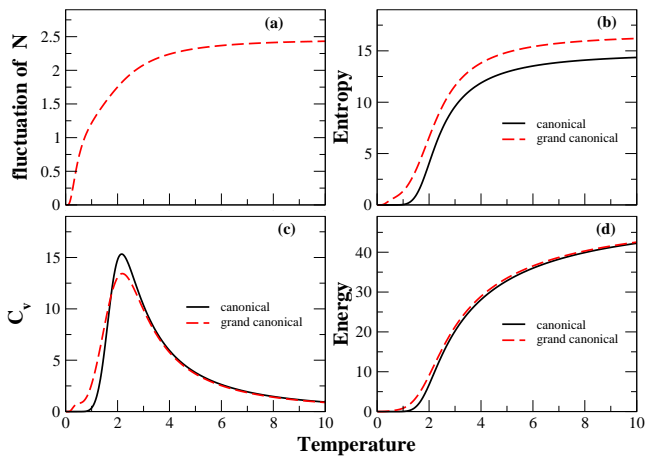}

\caption{(Color online) Thermodynamic properties of the ladder system with
12 levels, 12 particles, and $V$=1.00 are shown as a function of
temperature. The grand canonical ensemble is compared to the canonical.
(a) Fluctuation in the number of particles in the grand canonical
ensemble; (b) entropy; (c) specific heat; (d) excitation energy. \label{fGC}}
\end{figure}

\begin{figure}
\includegraphics[clip,width=8.5cm]{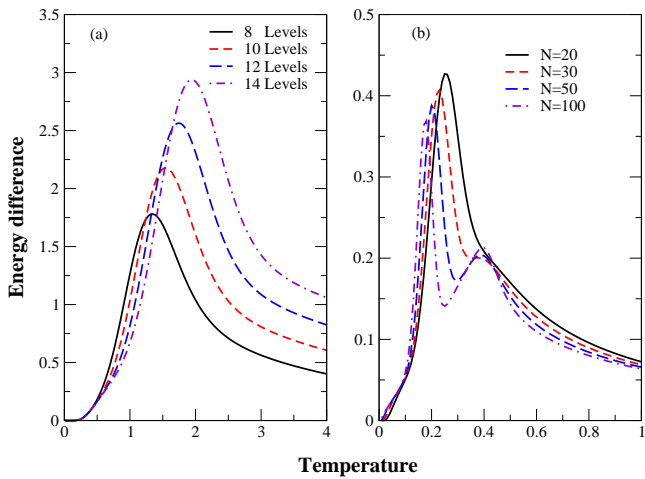}

\caption{(Color online) The excitation energy difference between canonical
and grand canonical statistical ensembles is shown as a function of
temperature. Left panel corresponds to a ladder system and two-level
model is on the right. \label{fig:C-GCenergy}}
\end{figure}

Various thermodynamic properties of the ladder system with 12 levels
and 12 particles obtained with the exact calculation of the grand
canonical partition function are shown in Fig. \ref{fGC}, the figure
also includes comparisons with the corresponding curves from the canonical
ensemble, where applicable. The fluctuations in the particle number
as a function of temperature are shown in Fig. \ref{fGC}(a). The
value of this quantity levels at about two particles, a similar uncertainty
on a level of one pair is present in the BCS theory. The particle
uncertainty relative to the system size $\sim2/N$ can be used to
estimate the quality of the grand canonical ensemble in applications
to particle-conserving mesoscopic systems. The results of comparisons
between canonical and grand canonical ensembles for the entropy, excitation
energy, and specific heat as a function of temperature are shown in
Fig. \ref{fGC}(b)-(d). The difference is quite small, and is consistent
with the level of error from the particle non-conservation. Further
comparison is shown in Fig. \ref{fig:C-GCenergy} where the energy
difference between canonical and grand canonical ensembles is plotted
as a function of temperature. In the picket-fence model the discrepancy
is noticeable, however it becomes relatively small in the two-level
case with a much larger number of particles. The difference peaks
exactly at the temperature of the phase transition (in both cases
$V=1$) where fluctuations are large. As seen in the two-level model
for a large number of particles this region becomes narrow. Although
for a ladder system the difference grows in the absolute scale, this
behavior is associated with the extreme smallness of the system and
the discrepancy per particle is still going to zero.

The zeros of the analytic continuation of the grand partition function
into the complex plain of chemical potential are of a separate interest.
We start by defining these points with the set of complex numbers
$\mu_{i}$ that for a certain temperature satisfy the equation ${\bf Z}(\beta,\mu_{i})=0.$
There are some features to be stressed here. The number of principal
roots $\mu_{i}$ is equal to the capacity of the fermion space $\Omega$.
The grand canonical partition function (\ref{eq:GCPart2}) is an $\Omega$-th
order polynomial in fugacity which leads to $\Omega$ roots in the
chemical potential that can be found with the standard numerical techniques
for polynomials. The methods discussed in the context of the canonical
partition function are also useful in this case. As the size of the
system grows the roots increase in number and may form branches that
can approach the real axis. This describes the mesoscopic phase transition
within the Yang-Lee picture \cite{YANG:1952,LEE:1952}. The accumulation
of roots near the real axis, similarly to the canonical ensemble discussed
earlier, represents a phase transition marked by the discontinuity
in a certain order derivative of the grand canonical partition function
with respect to chemical potential. This leads to the discontinuity
in the pressure-volume diagram \cite{Huang:1987} and in the thermodynamic
potential as a function of the particle number, namely condensation.
Based on the well known properties of the Bose gas the appearance
of such third order transition \cite{Green:1958} could be a good
evidence for the Bose-Einstein pair condensation. Whether with the
increased pairing strength or in a certain limit of temperature the
Cooper pairs become dynamically equivalent to bosons and form a condensate
and if there is a crossover region is an interesting and important
question \cite{Landau:1978,Koh:1997}.

Before addressing the results of this study we discuss some of the
expected features that can be inferred from the partition function
(\ref{eq:GCPartition}). Within the BCS approximation the integral
\eqref{eq:GCPartition} is given by the single saddle point value
\[
{\bf Z}_{BCS}={\bf Z}_{0}\exp(-\frac{2\beta\Delta^{2}}{G})\,\prod_{1>0}\left[\frac{\cosh\left(\frac{\beta}{2}e_{1}\right)}{\cosh\left(\frac{\beta\epsilon_{1}}{2}\right)}\right]^{4}.\]
 The Yang-Lee zeros of the above expression are zeros of the hyperbolic
cosine and for each single particle energy $\epsilon_{1}$ an infinite
series of roots labeled by integer $n$ can be obtained \begin{equation}
\mu=\epsilon_{1}\pm i\sqrt{\Delta^{2}+\pi^{2}\left(\frac{2n+1}{\beta}\right)^{2}}.\label{eq:murootsBCS}\end{equation}

\begin{figure}[ht]
\vskip 0.4 cm \includegraphics[clip,width=8.5cm]{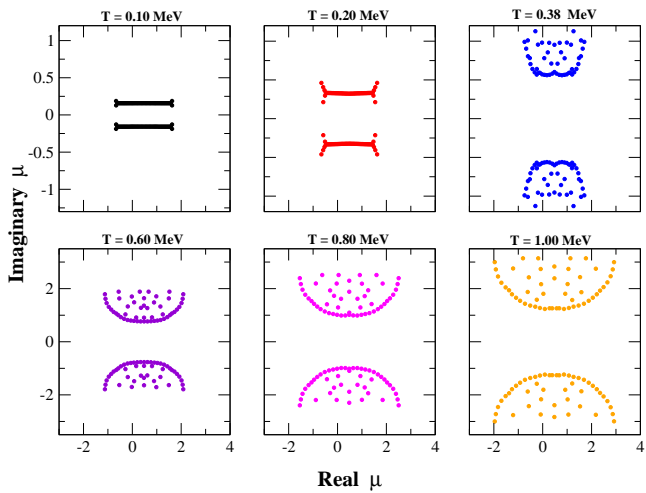}

\caption{(Color online) The distribution of zeros (DOZ) of grand canonical
ensemble in the complex chemical potential plane for a two levels
system with $V$=50 and $V$=1.00 for various temperatures indicated.
The $T_{c}=0.38$ for this system. \label{dozgcmu}}
\end{figure}

\begin{figure}[ht]
\vskip 0.4 cm \includegraphics[clip,width=8.5cm]{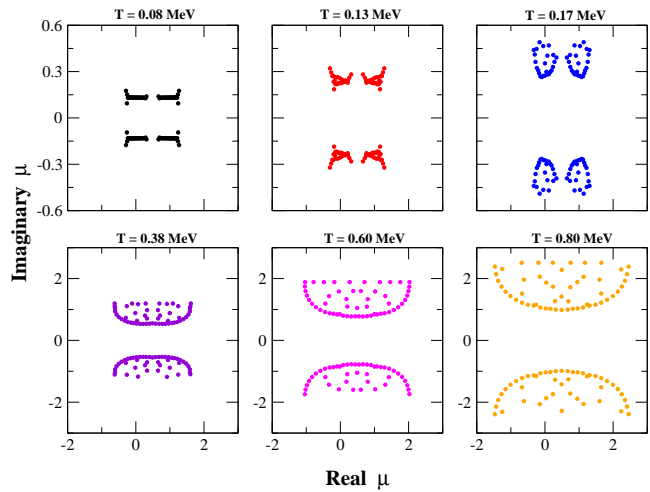}

\caption{(Color online) Same as Fig. \ref{dozgcmu}, except $V=0.5$ and corresponding
$T_{c}=0.17.$ \label{dozgcmuG05}}
\end{figure}

\begin{figure}[ht]
\vskip 0.4 cm \includegraphics[clip,width=8.5cm]{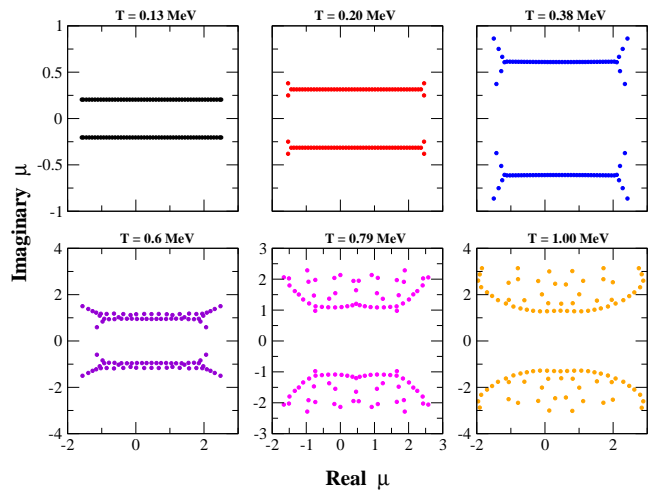}

\caption{(Color online) Same as Fig. \ref{dozgcmu}, except $V=2$ and thus
$T_{c}=0.8.$\label{dozgcmuG2}}
\end{figure}

The evolution of DOZ of grand canonical ensemble in the complex plane
of chemical potential for a fixed pairing strength G and various temperatures
is shown in Figs. \ref{dozgcmu}-\ref{dozgcmuG2}. In all of the plots
only the principal branch of roots with $n=0$, Eq. \eqref{eq:murootsBCS},
which is closest to the real axis is shown.

In Fig. \ref{dozgcmu} a somewhat high pairing strength $V=1$ is
selected so that at low temperature the system is well in the superconducting
phase. The resulting zeros are located along the horizontal line consistently
with Eq. (\ref{eq:murootsBCS}). As the temperature increases the
two lines of roots move apart deeper into the complex plane; this
trend is again in agreement with (\ref{eq:murootsBCS}), however the
overall behavior of the roots is no longer regular. The critical temperature
in this system (from the peak in heat capacity), is $T_{c}=0.38$
which coincides with a region where the behavior of DOZ changes. As
seen from the figure there are no branches of substantial significance
that cross real axis, indicating no phase transition.

The following Figs. \ref{dozgcmuG05} and \ref{dozgcmuG2} repeat
the same study with weaker and stronger pairing. The findings are
similar: at about critical temperature the DOZ changes from the two-line
distribution, reflecting the BCS limit, to a more scattered set of
roots moving away from the real axis at high temperature.

This exact calculation is consistent with the similar study \cite{Koh:1997}.
At temperatures below critical and with strong pairing, Fig. \ref{dozgcmuG2},
there are small symmetric branches of zeros that are directed toward
the real axis (although never reach it), they do not appear to result
in any transitional behavior and their significance is unclear. Our
models lack the explicit spatial degree of freedom and it is likely
that the BCS-BEC transition that reflects the change in the physical
size of the Cooper pairs is simply impossible here.

The grand canonical partition function (\ref{eq:GCPartition}) is
useful for understanding DOZ in the complex temperature plane, although
canonical and grand canonical ensembles are not fully equivalent.
Balian and Langer \cite{Balian:1963} have determined that the zeros
approach the real axis at an angle $\nu=\pi/4$ and their density
tends linearly to zero, $\alpha=1$, showing a second order transition.

\subsection{Microcanonical ensemble}

The microcanonical ensemble is often assumed to be the most physically
appropriate statistical treatment of a closed system. There is a number
of theoretical works as well as direct experimental studies of nuclear
thermodynamics in the microcanonical ensemble \cite{Guttormsen:2000,Guttormsen:2001,Guttormsen:2003,Schiller:2003}.
The density of states (DOS) $\rho$ is the primary element in the
approach. Regrettably, the formal definition given earlier \eqref{eq:denmicro}
is not appropriate per se, the density of states as well as the normalization
in the discrete spectrum requires some averaging energy interval.
For most of this study we chose not to implement a traditional binning
method substituting it by the propagator-type approach, where an artificially
inserted small width $\sigma$ (the same for all states) results in
the Lorentzian-type smoothing of every peak. The derivatives of the
DOS are then calculated based on the analytic derivatives of the Lorentzian
which provides an additional stability. With this procedure the DOS
$\rho(E)$ is obtained. The averaging width $\sigma$ is an artificial
parameter that introduces thermodynamic averaging, the results may
strongly depend on this parameter when it is smaller than the average
level spacing. This parameter is not necessarily a disadvantage, to
the contrary, it allows us to zoom at the energy scale of interest.
Within this work we select $\sigma=0.5-1.0$ in single-particle level
spacing units. This is most reasonable micro scale and can be compared
with the resolution scale of the canonical ensemble where energy fluctuations
are at about 10. The $\sigma$ interval versus level spacing can be
interpreted as the number of states needed to obtain a statistically
equilibrated value for observable quantities, in the limit of quantum
chaos a single state is sufficient \cite{RefWorks:519,Zelevinsky:1996},
on the other hand as discussed below pure pairing due to seniority
conservation is poorly equilibrated and many states must be included.
The latter fact influenced our choice of $\sigma$.

The entropy in the microcanonical ensemble is \begin{equation}
S(E)=\ln\rho(E),\end{equation}
 and the temperature can be defined as \begin{equation}
T(E)=\left(\frac{\partial S(E)}{\partial E}\right)^{-1},\label{eq:MCEtemperature}\end{equation}
which does not depend on the normalization that is used for the DOS.

\begin{figure}
\includegraphics[clip,width=8.5cm]{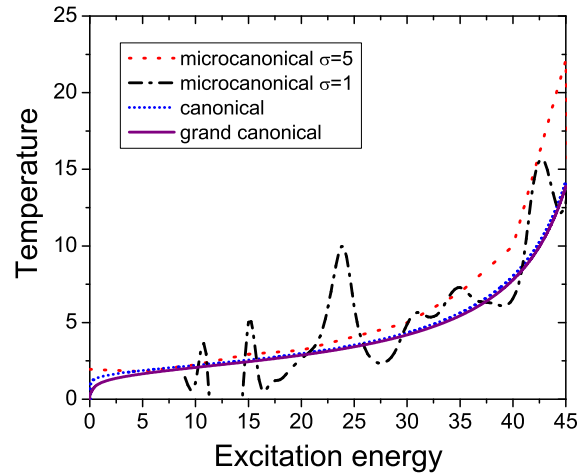}

\caption{(Color online) Temperature as a function of energy in three different
statistical treatments: canonical, grand canonical and microcanonical.
The ladder system with 12 levels, 12 particles and $V$=1.00 is used
for this study. The results for the microcanonical ensemble are plotted
with two different choices of energy window $\sigma=1$ and 5.\label{fMC}}
\end{figure}


%
\begin{figure}
\vskip 0.2 cm \includegraphics[clip,width=8.5cm]{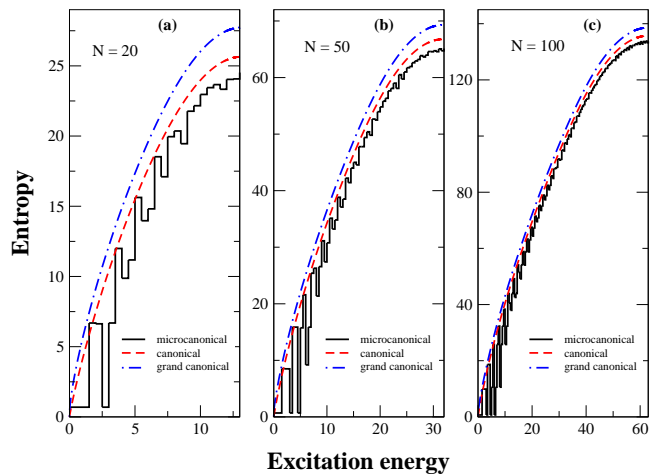}

\caption{(Color online) Comparison of entropy as a function of excitation
energy for the two-level system in three thermodynamic ensembles ($\sigma=0.5$
in microcanonical). Pairing strength $V$=1.00. (a) For N=20 particles;
(b) N=50 particles; (c) N=100 particles.\label{SMC2L}}
\end{figure}

In Fig. \ref{fMC} the temperature is shown as a function of excitation
energy for all three ensembles. The microcanonical curve with $\sigma=1$
shows several low-lying peaks that can be identified with the pair
breaking \cite{Schiller:2001,Guttormsen:2003,RefWorks:900}. The seniority
is a conserved quantum number for pure pairing interaction, nevertheless
these peaks survive in the presence of all interactions as was shown
in Ref. \cite{RefWorks:742,RefWorks:758,Guttormsen:2000}. The corresponding
oscillations in the heat capacity and especially the regions where
this quantity is negative can be associated with the paired to normal
phase transition which takes place in the pair-by-pair microscopically
fragmented process. The canonical and grand canonical ensembles due
to thermalization created by the heat bath smooth out these microscopic
features into a single phase transition, and importantly, the same
can be done in the microcanonical treatment by choosing a large averaging
interval $\sigma$. Already at $\sigma=5$, which is about half the
energy fluctuation in the canonical ensemble, the peaks in Fig. \ref{fMC}
disappear and microcanonical approach becomes similar to canonical
and grand canonical. 

The macroscopic limit of the microcanonical ensemble is considered
in Fig. \ref{SMC2L} where entropy as a function of excitation energy
in the two-level system for various $N$ is shown. The comparison
of microcanonical, canonical and grand canonical treatments indicates
that they become identical in the thermodynamic limit. The discrepancy
at high energy is related to the finite space where in the microcanonical
case the density of states becomes zero once the energy exceeds the
maximum possible value for the model space. The model space limitation
is a natural cut-off for all ensembles at high energy. 

In contrast to the canonical and grand canonical ensembles where thermalization
is provided by the external heat bath the thermalization is a serious
question in the microcanonical treatment \cite{Flambaum:1997,Flambaum2:1997}.
It has been noted in Refs. \cite{RefWorks:758,RefWorks:764,RefWorks:900}
that pairing interactions do not provide sufficient thermalization.
Particle-particle interactions of the pairing type only are not sufficient
to fully mix states and thermodynamically equilibrate the system.
Temperature determined microscopically (\ref{eq:MCEtemperature})
is inconsistent with the one that comes from the occupation numbers
of individual single-particle levels see Ref. \cite{RefWorks:758,RefWorks:764,RefWorks:900}.
This property of pairing makes the microcanonical treatment special.
The question of thermalization in systems with pure pairing is rather
academic; as it has been shown in \cite{RefWorks:900} and references
therein, at an arbitrary weak non-pairing interactions the equilibration
is immediately restored. The significant role of non-pairing interaction
was explored in recent work \cite{Horoi:2007}. The magnetic field
discussed below can also provide the needed thermalizing effect. The
sharp differences in the statistical approaches seen in Fig. \ref{fMC}
suggest to look for an alternative treatment and tracking of the transitional
behavior which would not introduce the heat bath, energy averaging,
or the particle number uncertainty but at the same time is statistically
equilibrated. The Invariant Correlational Entropy in the next section
provides a tool that satisfies this criteria.

\section{Invariant Correlational Entropy}

\begin{figure}[ht]
\begin{tabular}{cc}
\resizebox{60mm}{!}{\includegraphics[clip]{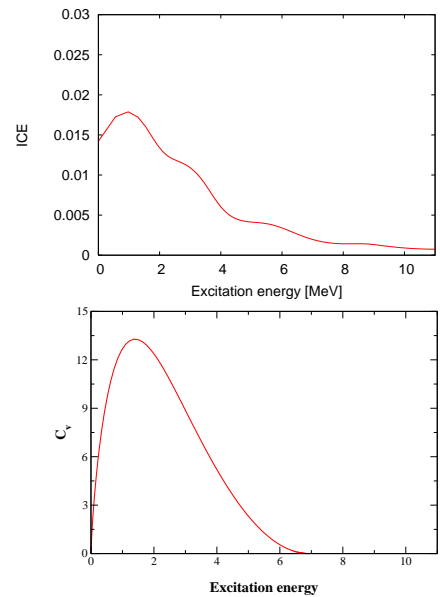}}

~~~&
\tabularnewline
\resizebox{51mm}{!}{\includegraphics[clip]{CE10}} &
\tabularnewline
\end{tabular}

\caption{(Color online) Invariant correlational entropy (ICE), upper panel;
and the specific heat in the canonical ensemble, lower panel, are
shown as a function of excitation energy for the two-level system
with N=10 particles and $V$=1.}

\label{ice} 
\end{figure}

The Invariant Correlational entropy (ICE) \cite{Sokolov:1998,Volya:2003su}
is a powerful method that allows phase transition features to be explored
on a quantum mechanical level. Expanding the formal definitions of
Sec. \ref{sec:Statistical-treatment}, the ICE for an individual eigenstate
$\alpha$ is computed by averaging the density matrix over the interval
of pairing strength \[
I^{\alpha}=-{\rm Tr}(\overline{\rho^{\alpha}}\ln\overline{\rho^{\alpha}}),\quad\overline{\rho_{k\, k'}^{\alpha}}=\overline{\langle k|\alpha\rangle\langle\alpha|k'\rangle},\]
 here $k$ is a basis state. The final result due to the trace operation
is basis independent. In Fig. \ref{ice} we show the invariant correlational
entropy for all states in the paired $N=10$ two-level system as a
function of the excitation energy of the corresponding state. The
ICE fluctuates from state to state and the curve shown has been smoothed.
The enhancement of the ICE in the region between 0 and 6 energy units
of excitation signals a transitional behavior. Indeed, a lower figure
that shows the specific heat as a function of energy for the same
system in the canonical ensemble reveals a consistent trend. The advantage
of the ICE is that unlike canonical (or grand canonical) ensemble
it needs no heat bath and maintains an exact particle conservation;
on the other hand, it is not prone to equilibration and thermalization
issues since those are established by the fluctuations of the pairing
strength.

\section{Pairing and Magnetic Field \label{sec:Magnetic-Properties}}

The presence of magnetic field is known to influence the physics of
the pairing state and the pairing phase transition. In this section
we extend our study by showing the changes to the results brought
by the presence of the field. The Hamiltonian to be considered here
is \begin{equation}
H_{B}=H-{\rm g}J{\bf \cdot B},\label{eq:HB}\end{equation}
 where $J$ is the angular momentum of the state and $B$ is the magnetic
field. Without loss of generality we choose units of the magnetic
field so that the gyromagnetic ratio g=1. The introduction of the
magnetic field does not require a new diagonalization of the Hamiltonian.
All eigenstates shift in energy according to their magnetic quantum
number $M$, with the quantization axis along the field ${\bf B}$.
The problem outlined with the Hamiltonian \eqref{eq:HB} is identical
to the cranking model of rotating nuclei where $H_{\omega}=H-J\cdot\omega$
with $\omega$ representing rotational frequency. Thus, a reader interested
in rotations should understand {}``magnetic field'' as a {}``rotational
frequency''. 

In the case of a single particle on one level the magnetization is
a textbook example\begin{equation}
\langle M\rangle=\frac{1}{2}\left[(2j+1)\coth\left(\frac{1}{2}(2j+1)x\right)-\coth\left(\frac{x}{2}\right)\right].\label{eq:MB}\end{equation}
 The spin fluctuation $\chi(\beta,B)=\langle M^{2}\rangle-\langle M\rangle^{2}$,
related to spin susceptibility, is given by \begin{equation}
\chi=\frac{1}{4}\left(\text{csch}^{2}\left(\frac{x}{2}\right)-(2j+1)^{2}\text{csch}^{2}\left(\left(j+\frac{1}{2}\right)x\right)\right).\label{eq:CHIB}\end{equation}
 Here $x=gB/T$. The generalization of these results to the cases
with many levels is straightforward.

In our study below we assume that the degeneracy in the single-particle
spectrum is due to the corresponding value of the angular momentum
$j=\omega-1/2$ for the two-level model and $j=1/2$ for all levels
in the ladder system. For each seniority we deduce the number of states
with certain angular momentum which in turn allows us to calculate
statistical partition functions. The analytically computed summations
over the magnetic quantum numbers such as in Eqs. \eqref{eq:CHIB}
and \eqref{eq:MB} speed up the calculations.

The destruction of the superconducting state occurs because of the
two somewhat related phenomena. A magnetic field causes the breaking
of superconducting pairs due to the lowering in energy of the spin
aligned states. The estimate for the critical field in this case can
be obtained by comparing the energy of the paired ground state with
the seniority $s=2$ aligned spin state with spin $J$. The latter
is by $2\Delta$ higher in energy at zero field and pairing becomes
unfavorable when magnetic field exceeds $B_{cr}=2\Delta/{\rm (gJ}).$
In our models, equations such as \eqref{eq:BCS-2l} or \eqref{eq:VBthermal}
can be used for an estimate. The second reason is the change in the
energy of the normal state reflecting the Pauli spin paramagnetism.
In our models (half-occupied for the two-level case) the field above
$B_{cr}=\Delta\epsilon/({\rm 2g}j)$ will promote the particle-hole
excitations across the gap between the single particle levels. It
has been suggested \cite{Maki:1966} and experimentally confirmed,
for example in Ref. \cite{Bianchi:2002}, that due to these phenomena
a sufficiently strong magnetic field could change the transition type
from second to first order. The situation can be quite complex in
mesoscopic systems where even without the field the classification
can be somewhat difficult.

\begin{figure}
\vskip 0.2 cm \includegraphics[clip,width=8.5cm]{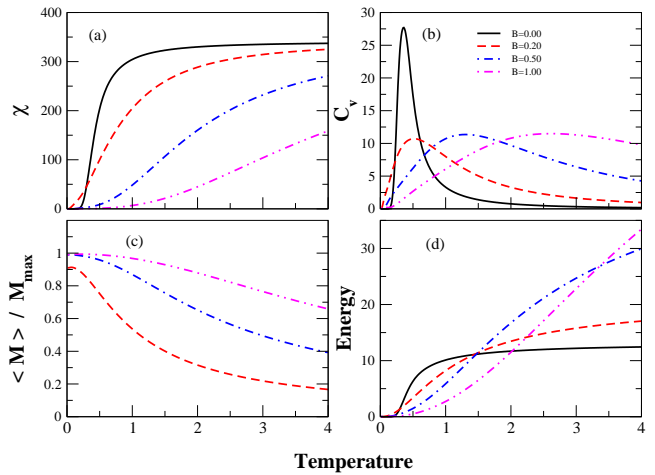}

\caption{(Color online) Thermodynamics of the two level system with $N$=20
and $V$=1.00 in the magnetic field. The (a) spin fluctuations $\chi$;
(b) specific heat; (c) magnetization (ratio to the maximum possible
value); and (d) energy are shown as functions of temperature for different
field strengths. For this model $T_{cr}=0.46$ at $B=0.$ \label{mag2L}}
\end{figure}

The features of a 20-particle half-occupied two-level system are shown
in Fig. \ref{mag2L} as a function of temperature for a set of different
values of magnetic fields. Except for $B=0$, for all curves in this
figure the field exceeds both of the above critical values ($B_{cr}\sim0.1$).
The behavior of the heat capacity illustrates the disappearance of
the phase transition for all field strength shown, except for $B=0$
where a sharp peak is present. The average magnetization Fig. \ref{mag2L}(c)
is exactly zero in the absence of field and for high fields starts
almost at the saturation. With increased temperature magnetization
drops down.

\begin{figure*}
\vskip 0.2 cm \includegraphics[width=15cm]{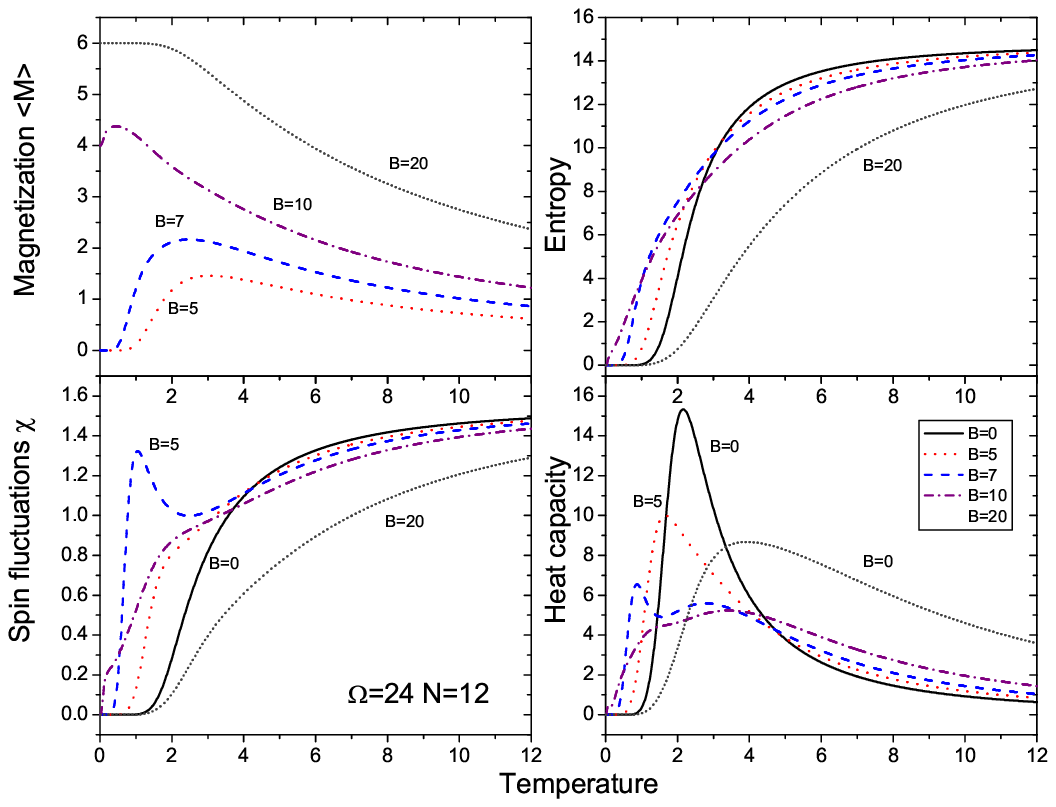}

\caption{(Color online) Thermodynamics within the canonical ensemble of a
12-level ladder system with $N=12$ particles in the magnetic field.
The pairing strength is $V=1$. The panels include plots of magnetization,
entropy, magnetic susceptibility and specific heat as a function of
temperature. The curves correspond to five different values of the
magnetic field B=0, 5, 7, 10 and 20. \label{fig:O12N12}}
\end{figure*}

\begin{figure*}
\vskip 0.2 cm \includegraphics[width=15cm]{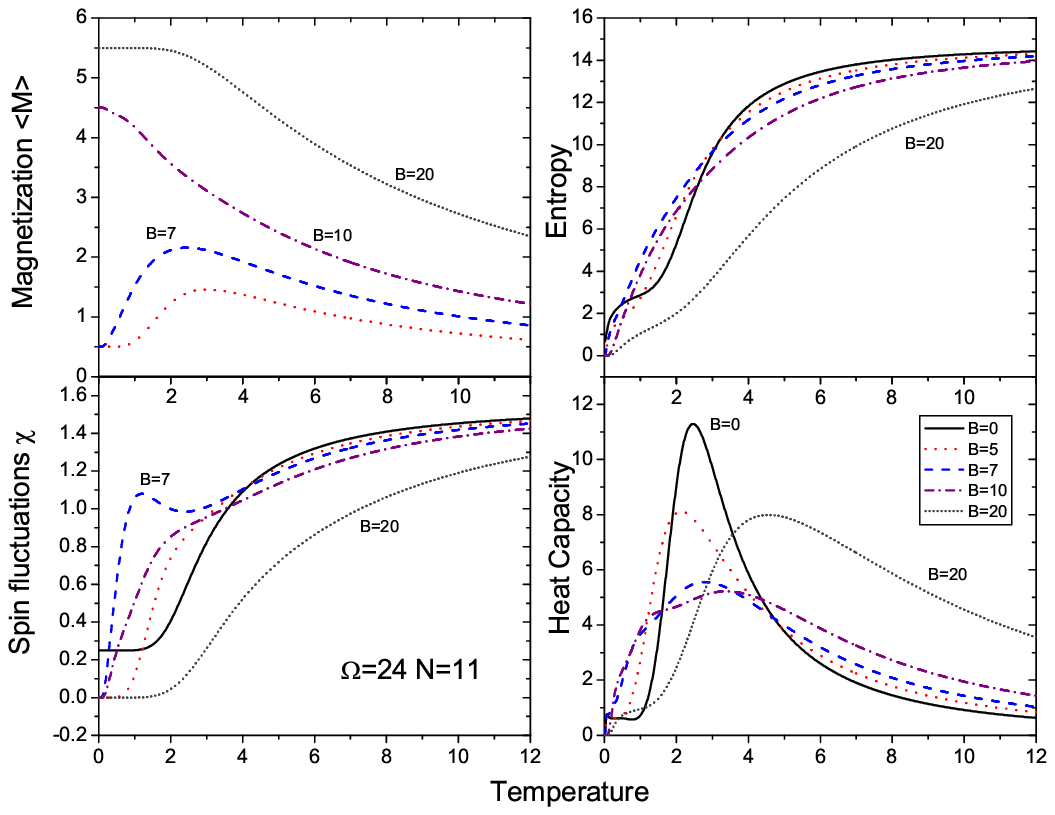}

\caption{(Color online) Thermodynamics within the canonical ensemble of a
12-level ladder system with 11 particles in the magnetic field. The
pairing strength is $V=1$. The panels include plots of magnetization,
entropy, magnetic susceptibility and specific heat as a function of
temperature. The curves correspond to five different values of the
magnetic field B=0, 5, 7, 10 and 20. \label{fig:O12N11}}
\end{figure*}

The set of fields below critical is shown in the next figure \eqref{lowmag2L}.
The behavior of the magnetic spin fluctuations is regular at $B=0$,
at weak fields this quantity exhibits a sharp peak at low temperatures,
for strong fields the regular behavior is again restored. The same
peak is present in the spin susceptibility curve which is only by
a factor of $\beta^{2}$ different from $\chi.$ The critical behavior
is associated with the corresponding behavior of the magnetization,
see Figs. \ref{fig:O12N12} and \ref{fig:O12N11}. The peak in the
heat capacity is reduced and shifted to lower temperature, showing
that at non-zero field a superconductor has lower critical temperature.

\begin{figure}
\vskip 0.2 cm \includegraphics[clip,width=8.5cm]{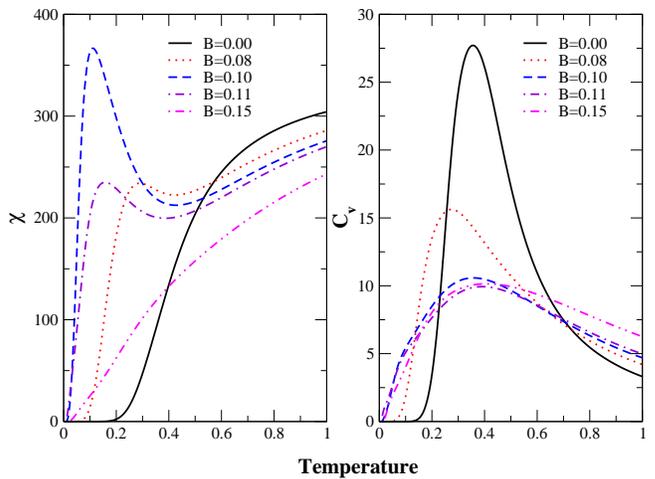}

\caption{(Color online) The same system as in Fig. \ref{mag2L} but concentrating
on the low magnetic fields. (a) spin fluctuations, (b) specific heat.
\label{lowmag2L}}
\end{figure}

\begin{figure}
\vskip 0.2 cm \includegraphics[width=8.5cm]{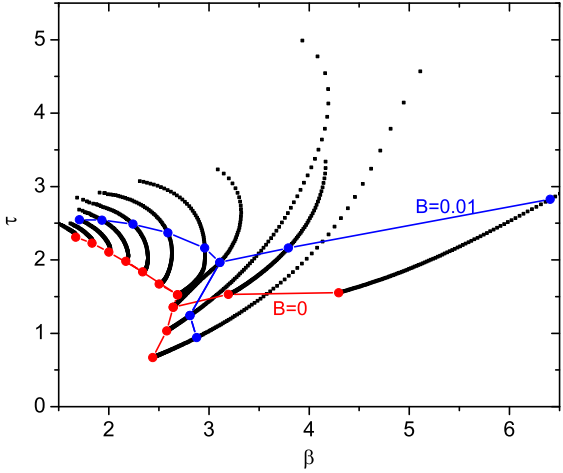}

\caption{(Color online) The distribution of zeros without and with the presence
of external magnetic field for two level system N=60 and G=1.00 .
\label{dozB}}
\end{figure}
\begin{figure}
\includegraphics[clip,width=8.5cm]{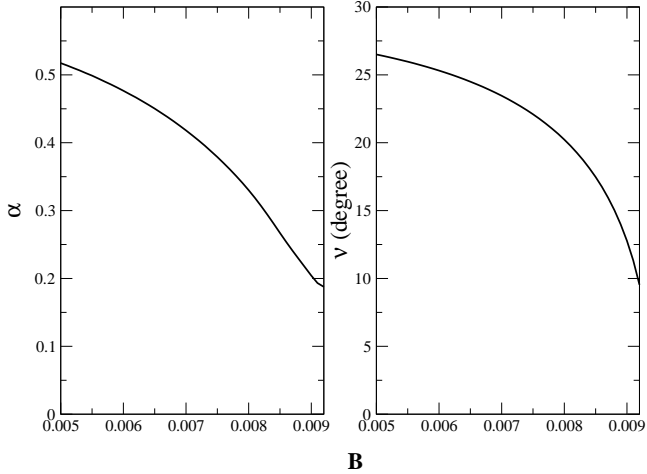}

\caption{The evolution of critical parameters $\alpha$ and $\nu$ as a function
of applied magnetic field $B$ in system with $N$=100 particles and
$V$=1.00 \label{fig:PTtypeB}}
\end{figure}

These results are quite robust. In Fig. \ref{fig:O12N12} we show
the same study repeated for the ladder model. The magnetization (upper
left panel in Fig. \ref{fig:O12N12}) is zero for no field, it shows
a peak in the region of the field strengths that corresponds to competition
between the paired state and magnetic orientation ($B=5$ and 7 curves),
and finally at strong fields and low temperature it saturates to $\langle M\rangle=6$,
the maximum alignment state. The spin fluctuations, shown on the left
lower panel, at zero temperature and no field, are consistent with
the typical results \cite{alhassid-2007}. The sharp peak at magnetic
fields below critical again reflects the transition associated with
magnetic alignment. The emergence of the peak that in thermodynamic
limit would become a discontinuity in the otherwise continuous curve
at $B=0$ can be interpreted as the change in the type of the phase
transition. The evolution of the transition point as a function of
temperature is seen in the heat capacity curve which again shows lowering
of the transition temperature with the increased field. The finding
can be summarized as the existence of two critical values for the
magnetic field first one that corresponds to the change in the nature
of the normal-to-superconducting transition; and at the second, higher
value of the field, the paired state is no longer supported. The change
in the nature of the phase transition is best seen in the spin-susceptibility
curve which at low fields has no peak and acquires a peak consistent
with the peak in heat capacity at higher values of the magnetic field.
We associate this behavior with the analogous situation in the macroscopic
superconductor where the second order phase transition becomes first
order in the non-zero magnetic field. 

Recently, sharp differences between the systems with odd and even
particle numbers have been discussed in the literature \cite{alhassid-2007,Schiller:2002av}.
We address this in Fig. \ref{fig:O12N11} where we show the same study
as in Fig. \ref{fig:O12N12} but with $N=11$ particles. The primary
difference between these results is that the ground state is degenerate
and both magnetization and entropy are non-zero at low temperature
and low field. Otherwise the results are almost identical. We conducted
similar calculations for a two-level model with $N=19$ particles
but decided not to show the results because the difference between
$N=20$ and $N=19$ is almost impossible to notice (except for the
entropy and magnetization at zero temperature and zero field).

The presence of the external magnetic field certainly has an effect
on the distribution of zeros in the complex temperature plane. This
evolution is explored in Fig. \ref{dozB} where the motion of roots
is traced as the magnetic field is increased in small increments.
The initial distribution of roots is connected with a line marked
by $B=0$. The second line connects the roots at $B=0.01.$ The gradual
rotation of the branch responsible for the phase transition is seen,
which eventually, at high fields, no longer orients the roots toward
the real axis. Based on a similar picture but for a large system with
$N=100$ particles that was studied earlier in Fig. \ref{fig:PTtypeB},
we show the change in the critical parameters associated with this
motion in the presence of the field. Interestingly, both $\alpha$
and angle $\nu$ are approaching zero which marks the change in the
phase transition type from second to first order. Unfortunately, the
zero $\alpha$ is not reached since the field strength becomes larger
then the critical (here $B_{cr}\sim0.01$) and the paired phase disappears.
The plot in Fig. \eqref{fig:PTtypeB} is ended at this point since
it is becomes impossible to identify a branch of roots relevant to
the phase transition.

\begin{figure}
\vskip 0.2 cm \includegraphics[width=8.5cm]{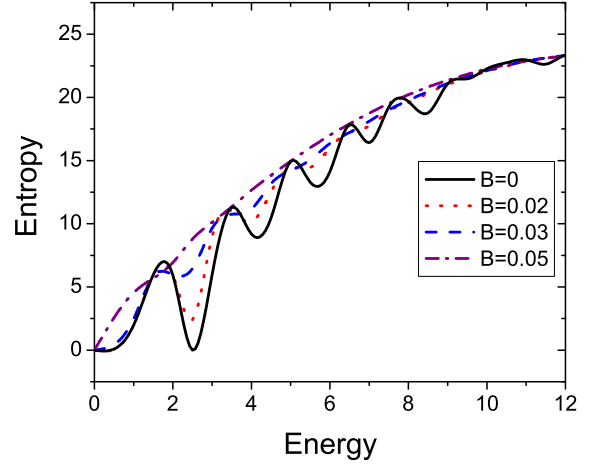}

\caption{(Color online) Entropy in the microcanonical picture is shown as
a function the excitation energy for several values of the magnetic
field below critical. The two-level system with 20 particles and $V=1$
pairing strength was used. \label{fig:entmat}}
\end{figure}

Finally we mention a thermalizing role of the magnetic field which
breaks high degeneracies of states; and as well as changing the phase
transition globally it reduces the number of individual pair breaking
transitions seen in the microcanonical treatment. In Fig. \ref{fig:entmat}
the entropy in the microcanonical ensemble is plotted as a function
of excitation energy for different strengths of the external magnetic
field. The number of peaks associated with the pair breaking is ten
at zero field and this number becomes smaller as the magnetic field
gets stronger, simply because of the pair alignment. Thermal equilibration
and the equivalence of ensembles are seen in the following Fig. \ref{fig:TvsEB}.
In contrast to the $B=0$ case in Fig. \ref{fMC}, the difference
between canonical and microcanonical ensembles disappears at the magnetic
field near critical.

\begin{figure}
\vskip 0.2 cm \includegraphics[width=8.5cm]{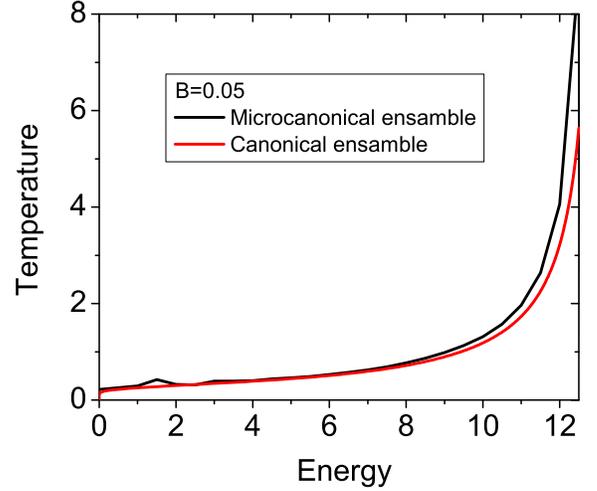}

\caption{(Color online) The same system as in Fig. \ref{fig:entmat}, the
temperature as a function of energy is compared in microcanonical
and canonical ensembles at the magnetic field strength $B=0.05$.
The difference at high energy is due to a finite model space. \label{fig:TvsEB}}
\end{figure}

\section{Summary Conclusions and Outlook}

Using an exact solution to the pairing problem in a picket-fence and
two-level systems we addressed different views on the pairing phase
transition, pair breaking, thermalization, behavior in the magnetic
field or, equivalently, rotation in the framework of the cranking
model. We present a comprehensive study analyzing paired systems with
various tools and methods ranging from the BCS treatment to different
thermodynamic approaches, including invariant correlational entropy
and zeros of the partition functions in the complex plane of temperature
and chemical potential.

We found the microcanonical, canonical, and grand canonical thermodynamic
approaches to be different when applied to small systems, although
as expected they are consistent in the macroscopic limit. The grand
canonical and canonical treatment are surprisingly close to each other
even in the cases with a relatively small particle number. We find
the microcanonical treatment to be the most adequate approach for
closed small systems, it allows for both global and relatively local,
in terms of the energy scale, consideration of the statistical properties.
The averaging energy window needed for determination of the density
of states gives a broad freedom to the microcanonical approach. The
corresponding energy fluctuation in canonical and grand canonical
treatments is too large in small systems and smooths out many significant
statistical features such as individual pair breaking observed in
experiment \cite{Guttormsen:2000}. In thermally equilibrated systems
the energy window can be as small as few times the level spacing since
in the full quantum chaos an individual state is a representative
of the surrounding statistical properties \cite{RefWorks:519}. The
idealization of interaction limiting them to pairing only represents
a dangerous problem: the pairing forces exclusively cannot establish
full equilibration, this necessitates a larger thermodynamic energy
window.

The Invariant Correlational Entropy that relies on fluctuations in
the pairing strength as a source of equilibration appears to be a
powerful statistical tool, capable of exploring most of the features
inherent separately to microcanonical, canonical and grand canonical
ensembles. This tool is particularly important in identifying phase
transition regions in mesoscopic systems. 

Turning to the properties of the phase transition we found as mentioned
above the microcanonical treatment to be distinct from other ensembles.
In thermodynamic limit, however, all treatments agree. In the further
study of phase transitions we discussed the distribution of zeros
of the canonical partition function in the complex temperature plane.
We developed and implemented a numerical method for counting and finding
all of the complex roots in a given region. In agreement with the
earlier findings \cite{Schiller:2002av,Dean:2002zx,Ipsen:2003} we
observe branches of roots and study the properties of the one that
approaches the real axis. The behavior of the roots is consistent
with the second order phase transition as classified in \cite{GROSSMAN.S:1968,GROSSMAN.S:1969}
and confirms similar macroscopic results \cite{Balian:1963}.

The recent interest to the crossover region between superconducting
pairing and Bose-Einstein condensation of pairs prompted us to consider
the potential condensation by looking for zeros of chemical potential
in the grand canonical partition function. The presence of such zeros
near real axis would hint on the condensation phenomenon. We did not
find significant branches of roots evolving toward the real axis,
and no critical behavior was observed in thermodynamics. It is likely
that our model with no explicit spatial degrees of freedom is not
appropriate for these questions. 

The last chapter of our work is devoted to interesting study of the
mesoscopic phase transition in the presence of magnetic field. It
is fully equivalent to rotations within cranking model. We found that
there is a resemblance between observed mesoscopic properties and
those known in the macroscopic physics of superconductors. At low
field the normal and superconducting phases are separated by the second
order phase transition. In the next region of higher magnetic field
the normal and superconducting phases are separated by the transition
of a different nature associated with a simultaneous peak in spin
susceptibility end enhanced spin fluctuations. Finally, at even higher
fields a superconducting state is not supported at all. We conjecture
that this behavior is a mesoscopic manifestation of the second to
first order change in the transition type known in the thermodynamic
limit. We also traced the evolution of zeros in the canonical partition
function as a function of magnetic field. We found that the classification
of transition type as suggested in Ref. \cite{Borrmann:2000} is consistent
with the above argument.

\section*{Acknowledgments}

The authors acknowledge support from the U. S. Department of Energy,
grant DE-FG02-92ER40750. We are grateful to V. Zelevinsky for collaboration
and invaluable advice on many topics presented in this work. We wish
to think A. Schiller, T. D{\o}ssing and P. Ipsen for useful comments
and references. 


\end{document}